\documentclass[fleqn,usenatbib]{mnras}

\usepackage[T1]{fontenc}
\usepackage{ae,aecompl}

\usepackage{graphicx}	
\usepackage{amsmath}	
\usepackage{amssymb}	
\usepackage{soul}
\usepackage{longtable}
\usepackage{caption}

\graphicspath{ {img/} } 

\title[UFDs and GCs in an evolving Galactic potential]
{The orbital evolution of UFDs and GCs in an evolving Galactic potential}

\author[B. M. Armstrong et al.]{
Benjamin M. Armstrong,$^{1}$\thanks{E-mail: benjamin.armstrong@icrar.org}
Kenji Bekki,$^{1}$
and Aaron D. Ludlow$^{1}$
\\
  $^{1}${International Centre for Radio Astronomy Research, University of Western Australia, 35 Stirling Highway, Crawley,}\\
  {Western Australia, 6009, Australia}\\
}

\date{Accepted XXX. Received YYY; in original form ZZZ}

\pubyear{2020}

\begin{document}
\label{firstpage}
\pagerange{\pageref{firstpage}--\pageref{lastpage}}
\maketitle

\begin{abstract}
We use the second \textit{Gaia} data release to investigate the kinematics of 
17 ultra-faint dwarf galaxies (UFDs) and 154 globular clusters (GCs) in the Milky Way,
focusing on the differences between static and 
evolving models of the Galactic potential. An evolving potential modifies a satellite's
orbit relative to its static equivalent, though the difference is small compared to existing uncertainties on orbital parameters.
We find that the UFD Bo\"otes II is likely on its first passage around the Milky Way.
Depending on the assumed mass of the Milky Way, the UFDs Triangulum II, Hydrus I, Coma Berenices, Draco II,
and Ursa Major II, as well as the GC Pyxis, may also be on first infall 
so may be useful for constraining the mass of the Galaxy. We identify a clear kinematic 
distinction between metal-rich (${\rm [Fe/H]}>-1.1$) and metal-poor GCs
(${\rm [Fe/H]}\leq-1.1$). 
Although most metal-rich clusters occupy predominately prograde orbits, with low 
eccentricities ($e\approx 0.35$) and similar specific angular momenta and orbital planes as the Galactic disc, 
7 show potentially retrograde orbits, the origin of which is unclear.
Metal-poor clusters have more diverse orbits, higher eccentricities ($e\approx 0.65$), and half 
have orbital planes offset from the disc by $60$ to $120$ degrees. The UFDs have similar $\theta$ and $\phi$ to the metal-poor GCs, suggesting 
a similar origin. We provide a catalogue of orbital parameters for UFDs and GCs for two different Galaxy masses 
and their observational uncertainties.

\end{abstract}




\section{Introduction}

The kinematics of Milky Way (MW) satellites provide important
clues for understanding the origin and evolution of the Local Group. 
Studying the kinematics  of the smallest MW satellites has
historically proven challenging because of the limited number of stars 
for which 6D phase-space data were available, the difficulty of measuring 
the mass of the Galaxy \citep[e.g. ][]{vanderMarel12, BlandHawthorn16}, 
and the unclear history of the Magellanic system 
\citep[e.g. ][]{Diaz12, Besla12, Kallivayalil13, Armstrong18}.
While many previous studies we able 
to track the orbits of the largest MW satellites---the Large 
(LMC) and Small Magellanic Clouds (SMC)
\citep{Murai80, Lin82, Gardiner94, Heller94, Gardiner96}---improvements to
observational techniques have recently revealed the proper motions of many of the MWs
dwarf galaxies and star clusters.

Recently, kinematic studies have undergone a revolution.
Where previous studies had to work with a limited sample of satellites with 
reliable proper motion measurements 
\citep[e.g. ][]{Cudworth93, Dinescu99, Dinescu00, Anderson03, Kalirai07, Eadie15},
the second data release of the \textit{Gaia} mission \citep[DR2; ][]{GaiaDR2} provided high-quality 
proper motions for more than 1.3 billion MW stars. This has allowed studies to probe the dynamical history 
of the Galaxy in greater detail than ever before \citep[e.g.][]{Baumgardt18, Fritz18, Kallivayalil18, Massari18, Simon18, Vasiliev19, Budanova19, Erkal19, Massari19, Pace19, Piatti19, Piatti19b, Watkins19, Patel20}.

Tracing the orbit of a MW satellite requires a model for the Galactic potential,
which is often---though not always---assumed to be static.
For example, the long-term orbital evolution in an evolving potential has previously 
been studied for the LMC \citep{Zhang12}, 
for many globular clusters \citep{Haghi15}, and most recently for dwarf 
spheroidal and ultra-faint dwarf galaxies (UFDs; \citealt{Miyoshi20}).
Nevertheless, little is known about how their orbits differ between a static and evolving 
potentials, nor how they compare to the existing uncertainties on satellite trajectories.
We assess these issues in this paper, using two tracer populations: UFDs and GCs.

UFDs are poorly understood; their faintness makes detection challenging, impeding their 
study. Recent surveys, such as \textit{Gaia} \citep{GaiaDR1} 
and the \textit{Dark Energy Survey} \citep{DES}, capable of detecting these 
systems have lead to a surge of interest, and our understanding of these objects 
and their orbital histories has improved substantially over the past few years
\citep[e.g.][]{Fritz18, Kallivayalil18, Massari18, Simon18, Erkal19, Pace19, Patel20}.
Yet outstanding questions remain. For example, which of the UFDs are associated with, 
and so have accreted alongside, the LMC and SMC 
\citep{Bekki08, D'Onghia08, Drlica15, Jethwa16}?
Likely candidates are Carina II, Carina III, Horologium I, Hydrus I, and Phoenix I. Reticulum II 
and Tucana II are other possibilities 
\citep{Deason15, Sales17, Kallivayalil18, Simon18, Erkal19, Patel20}. 
Sculptor I, Tucana III, Segue 1 are also thought to have undergone recent interactions 
with the LMC and SMC \citep{Patel20}.

Although a limited number of UFDs have been detected, the diversity between their individual 
orbits is already apparent. \cite{Simon18} studied the orbits of 17 UFDs in a 
low-mass MW potential ($M_{\rm vir}=0.8\times10^{12} {\rm M_{\odot}}$) and calculated
eccentricities ranging from 0.19 (Willman 1) to 0.96 (Bo\"otes II), with apocentres 
ranging from 53 to 1746 kpc, respectively.

GCs act as fossil records of the chemical and dynamical evolution of galaxies.
They are important tools for studying the structure the Galactic potential
\citep[e.g.][]{Wakamatsu81, Armandroff89, Bergh93, Dinescu99, Dinescu03, Bellazzini04, Watkins19}
and understanding how it affects stellar orbits
\citep[e.g.][]{Ostriker72, Spitzer73, Aguilar88, Chernoff90, Hut92, Capaccioli93}.
Cosmological simulations suggest that galaxies form through a combination of \textit{in-situ} 
star formation and accretion \citep{Oser10, Cook16, Rodriguez16}, indicating that they 
may also contain both \textit{in-situ} and accreted GCs \citep{Forbes18}. 
Because of their proximity, Galactic GCs can be resolved down to individual stars, 
making them ideal for understanding how \textit{in-situ} and accreted GCs differ and, 
through their orbits, revealing the merging histories of the galactic progenitors that 
formed the MW \citep{Searle78, Beasley02}.

The orbital histories of Galactic GCs are poorly constrained, with estimates of the fraction of 
accreted GCs reaching as high as 87 per cent \citep{Mackey04, Mackey05, Forbes10, Leaman13, Boylan-Kolchin17}.
Clear differences between GCs can be seen in both their colour and direction of motion. 
Nearly every massive galaxy studied with sufficient resolution shows a clear bimodal 
distribution of cluster metallicities \citep{Brodie06}. In the MW, the GCs can be approximately divided 
into a metal-poor population and a metal-rich population below and above [Fe/H] = $-1.1$, respectively
\citep[2010 edition]{Zinn85, Reed94, Harris96}. This could be caused by mergers in the early Universe 
\citep{Ashman92, BC02, Bekki02, Beasley02, Tonini13}, late mergers \citep{Gnedin10, Renaud17}, 
or a non-linear colour-metallicity relation \citep{Cantiello12}. Another argument comes from the 
significant population of retrograde clusters in the Galaxy. GCs formed \textit{in-situ} 
should be moving in the same direction around the Galaxy as the material they formed from, so 
should occupy prograde orbits. Accreted GCs have no such restriction; whether they are prograde 
or retrograde depends the nature of their specific accretion event.
Because about 25 per cent of the Galactic GCs have retrograde 
motion, studies have argued that up to half of the MW GCs may have been accreted
\citep{Mackey04, Mackey05, Forbes10, Vasiliev19, Kruijssen19,  Massari19, Piatti19}.

In this paper we investigate the orbits of the 17 UFDs presented in \citet{Simon18}, 
and the 154 GCs from the \citet{Baumgardt18} catalogue, both of which are based upon \textit{Gaia} DR2. 
In Section 2 we discuss our model for the MW potential and describe how we calculate
the trajectories for each satellite. Section 3 contains our results: we
present kinematic properties for each satellite and explore the differences between models. 
We discuss these results and summarise our conclusions in section 4.

\begin{table}
	\centering
	\caption{Description of the bulge and disc components of each model, as well as whether the potential 
	is static or evolving. The halo mass, $M_{\rm h}$, take values in the range
	$M_{\rm h}=0.6$--$2.4\times10^{12} {\rm M}_{\odot}$.}
	\label{Models}
	\begin{tabular}{lccc}
		\hline
        Model & $M_{\rm d}$ (${\rm M}_{\odot}$) & $M_{\rm b}$ (${\rm M}_{\odot}$) & Evolving\\
		\hline
        M1 & $10^{11}$ & $10^{10}$ & Yes\\
        M2 & $10^{11}$ & $10^{10}$ & No\\
        M3 & $3\times10^{10}$ & $3\times10^{9}$ & Yes\\
        M4 & $0$ & $0$ & Yes\\
		\hline
	\end{tabular}
\end{table}

\section{Preliminaries}

\subsection{Observational data}

For the UFDs we used the 
17 dwarfs presented in \citet{Simon18}, which provides the necessary kinematic parameters 
and their associated error, and determines proper motions from \textit{Gaia} DR2 using 
spectroscopic membership. We chose not to use the rest of the 39 dwarf galaxies 
presented in \citet{Fritz18} so that we could focus on just UFDs and how they evolve, and did 
not include the 7 UFDs presented in \citet{Pace19} because we did not have the radial velocity 
data necessary to properly analyse their orbits. 

For the GCs we used the \citet{Baumgardt18} catalogue of 154 clusters. The results from 
\citet{Baumgardt18} are in good agreement with similar studies \citep{Vasiliev19, Gaia}, 
and additionally provide radial velocities for each GC, which are vital for calculating 
orbits. \citet{Baumgardt18} combines their own distance measurements based on fitting 
velocity dispersion profiles to N-body models with existing catalogues of GC distances 
\citep[2010 edition]{Harris96} to determine the distance to each cluster and the distance 
uncertainty for 53 of the clusters. For the other 101 GCs with no measurements of 
distance uncertainty, we introduced an error margain of 10 per cent. While this is 
significantly higher than the errors presented in \citet{Baumgardt18}, potentially 
biasing orbits to favour an accretion scenario, we have elected to err on the side of 
caution. Furthermore, it is very possible that some GCs will still fall outside of 
this range.

\subsection{A model for the Milky Way}

We construct the Galactic potential using three components: a disc, a bulge, and a dark matter halo. 
The gravitational potential of the disc is represented by a Miyamoto--Nagai profile \citep{Miyamoto1975},
\begin{equation}
{\phi}(R,z)=-\frac{G M_{\rm d}}{\sqrt{R^2+(a+\sqrt{z^2+b^2})^2}},
\end{equation}
where $G$ is Newton's constant, $R$ is the Galactocentric distance in the plane of the disc, 
$z$ is the height above the disc, $M_{\rm d}$ is disc mass, and $a$ and $b$ are the scale length
and scale height of the disc, respectively. We adopt the values $a = 4$ kpc, and $b = 0.3$ kpc.
For the bulge we adopt a Hernquist potential \citep{Hernquist1990},
\begin{equation}
{\phi}(r)=-\frac{G M_{\rm b}}{r+c},
\end{equation}
where $r$ is the Galactocentric distance, $M_{\rm b}$ the bulge mass and $c$ the scale radius.
We adopt the value $c = 0.6$ kpc.
The scale parameters for the disc and bulge are comparable to those used in similar models of the 
Galactic potential
\citep[e.g.][]{Diaz12, Yozin12, Bajkova17, Errani17, Erkal18, Secco18, Erkal19}.

For the dark matter halo we adopt a Navarro--Frenk--White profile \citep[][hereafter NFW]{NFW1996},
\begin{equation}
{\phi}(r)=-\frac{G M_{\rm h}}{r} \ln \, \biggr(1+\frac{r}{r_{\rm s}}\biggl),
\end{equation}
where $M_{\rm h}$ is the halo mass and $r_{\rm s}$ its scale radius. We identify $M_{\rm h}$ with
the traditional virial mass, ${\rm M_{200}}$, which is the mass enclosed by a sphere of radius $r_{200}$
corresponding to a mean density $200\times\rho_{\rm crit}$ ($\rho_{\rm crit}=3{\rm H^2}/8\pi G$ is the critical
density for a closed universe). The halo's concentration is defined $c=r_{200}/r_{\rm s}$.

The structural parameters of a galaxy can be related to its mass accretion history (MAH; e.g.
\citealt{Ludlow14}). We model the halo using the empirical prescription present in \citet{COMMAHa, COMMAHb, COMMAHc}
(hereafter referred to as COMMAH), which describes time evolution of the {\em average} virial mass and concentration
for haloes of a given present-day mass, ${\rm M_{200}}(z=0)$. We use COMMAH to predict the evolution of the halo's
structural properties from the present-day to a look-back time of $t=13\,{\rm Gyr}$. Since COMMAH only models
the halo, the disc and bulge are evolved according to a simple prescription (eq.~\ref{eq:gas_evol}, below). 
We also investigate static models where the masses of each component---and hence the associated 
gravitational potential---are kept fixed to the present-day values.

We emphasize that COMMAH predicts the {\em smooth averaged} MAHs. Realistic cold dark matter 
haloes will instead experience bursts of activity as a result of occasional mergers.
\citet{Evans20} showed that an unusual assembly history is required to form a MW-like halo with 
a LMC-like counterpart, with a lower halo mass at earlier times. A more accurate handling of the MAH 
of the Galaxy would need to factor in its individual quirks.

We do not account for the possible contraction of the dark matter halo due to the presence of a stellar disc 
\citep[see e.g.][]{Schaller15, Zhu16, Dutton16, Cautun20}. While this may impact the orbits of the innermost GCs and UFDs, 
the effects are similar to adopting a larger disc mass or more concentrated dark matter halo \citep{Cautun20}. 
We instead adopt a relatively broad range of disc masses (Table~\ref{Models}) 
and avoid the uncertainties inherent to modelling halo contraction.

\begin{figure}
	\includegraphics[width=\columnwidth]{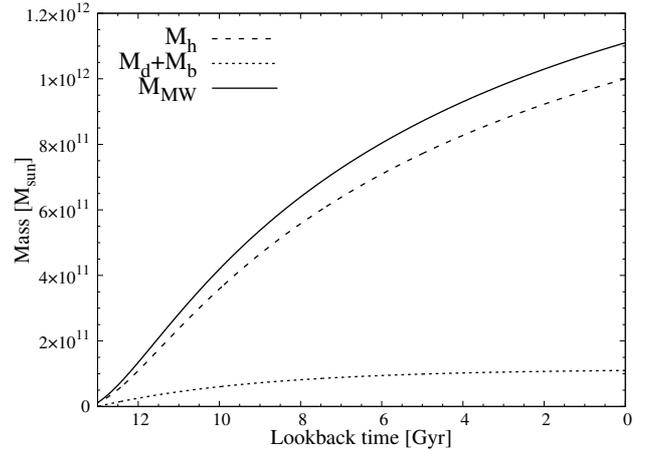}
	\vspace{-5mm}
    \caption{
    Halo and gas evolution of a sample potential with $M_{\rm h}=1.0\times10^{12}{\rm M}_{\odot}$, 
    $M_{\rm d}=1.0\times10^{11}{\rm M}_{\odot}$, and $M_{\rm b}=1.0\times10^{10}{\rm M}_{\odot}$.
    Each of the four model tests 10 different values for the halo mass. The disc and bulge masses  
    vary between models but are consistent within. 
    }
    \label{mass}
\end{figure}

In addition to the halo, we need to evolve the disc and bulge components of the potential. 
Because the mass of the halo is so dominant, the way we handle the evolution of the other 
components is less impactful. We evolve the bulge and disc using the simple prescription of
\begin{equation}
    \dot{M} \propto \exp{\biggr(-\frac{t}{4~{\rm Gyr}}\biggl)}.
\label{eq:gas_evol}
\end{equation}
The difference between this and a more sophisticated method is small compared to the case 
in which no evolution occurs, which is the primary focus of this study. The effect of this 
is shown in Fig.~\ref{mass}.

We investigate four models with different $M_{\rm d}$, $M_{\rm b}$, and with static 
and evolving potentials. The differences are summarised in Table~\ref{Models}. Model M1 uses 
disc and bulge components 
of $M_d=10^{11} {\rm M}_{\odot}$ and $M_b=10^{10} {\rm M}_{\odot}$, respectively,
and a Galactic potential that evolves according to COMMAH and eq.~\ref{eq:gas_evol}. 
M2 is similar to M1, but has a static potential. M3 has an evolving potential and smaller 
disc and bulge components of $3\times10^{10} {\rm M}_{\odot}$ and $3\times10^{9} {\rm M}_{\odot}$, respectively.
Finally, M4 has no disc or bulge components and consists of only an evolving dark matter halo.

\subsubsection{The mass of the Milky Way }

Within each model we test a range of present-day halo masses, from $M_{\rm h}=0.6\times10^{12} {\rm M}_{\odot}$ 
to $2.4\times10^{12} {\rm M}_{\odot}$, divided into 10 evenly-spaced intervals. 
This relatively broad mass range covers that presented by \citet{Watkins19} 
($M_{\rm MW} = 1.54^{+0.75}_{-0.44}\times10^{12} {\rm M}_{\odot}$), but has a lower limit roughly 50 per cent 
smaller to accommodate potential overestimation in MW mass estimates that do not account for the Magellanic Clouds
\citep{Erkal20}. We stress, however, that more precise measurements of the Milky Way's mass exist 
\citep[e.g.][]{McMillan17, Callingham19, Cautun20, Fritz20, Li20b}. A comparison of mass estimates performed by \citet{Wang20}
suggested that $M_{\rm MW} \approx 1.2\times10^{12} {\rm M}_{\odot}$, though different methods yield systematically
different results. We adopt a broad MW mass range to fully capture published mass estimates of the MW.
It also simplifies the comparison between our results and those of previous studies, such as \citet{Simon18}, 
which uses a MW mass of $M_{\rm MW} = 0.9\times10^{12} {\rm M}_{\odot}$.
While this method does mean that a given halo mass 
corresponds to slightly different {\em total} masses between models (due to the different $M_{\rm d}$ 
and $M_{\rm b}$) the impact is expected to be small since the total mass is dominated by the halo
in all cases.

\subsubsection{The Magellanic Clouds}

We have elected not to include the additional potential of the LMC and SMC.
The orbital history of the Magellanic Clouds is still a topic of some debate, 
so we cannot easily include a robust treatment of their gravitational potentials. 
A full exploration of the parameter space of the Magellanic Clouds would vastly increase 
the complexity of this work. 
To ensure that our study can focus primarily on the effect of mass accretion on orbits, we have 
used only the Galaxy's gravitational potential. While is unlikely to affect the orbits of the GCs, 
which are too closely tied to the MW, and distant UFDs, which are unlikely to have ever had a close 
approach with the LMC, the results for the UFDs in \citet{Simon18} that are associated with the LMC 
(Carina II, Carina III, Horologium I, Hydrus I, Reticulum II, Tucana II) or have had a close 
interaction with it (Tucana III, Segue 1) are unlikely to reflect these objects' true trajectories. 
In spite of this, the orbits of these UFDs can still provide valuable insight into how orbits in 
general can change in the presence of an evolving potential. No such problem is expected for the other nine UFDs. 

\subsubsection{Dynamical Friction}

The effects of dynamical friction on the orbits of close MW satellites is an important factor controlling 
their long-term evolution. However, this is primarily relevant for massive, close dwarfs. With 
the exception of Tucana III, all studied UFDs are too distant for dynamical friction to have much effect 
\citep{Simon18}, and the GCs are too low-mass to be strongly affected.
So that we can focus purely on the effect of the evolving potential, 
we have neglected dynamical friction in our model. This means that the evolution of 
the potential is the only thing that can change an orbit over time, ensuring that M2 is truly static. 
This is unrealistic for the GCs and Tuc II, the closest UFD, but it will allow us to separate out the results 
from the potential evolution and with clarity comment on whether it is worth including in future studies.

\subsection{Calculating orbital histories}

\begin{figure}
	\includegraphics[width=\columnwidth]{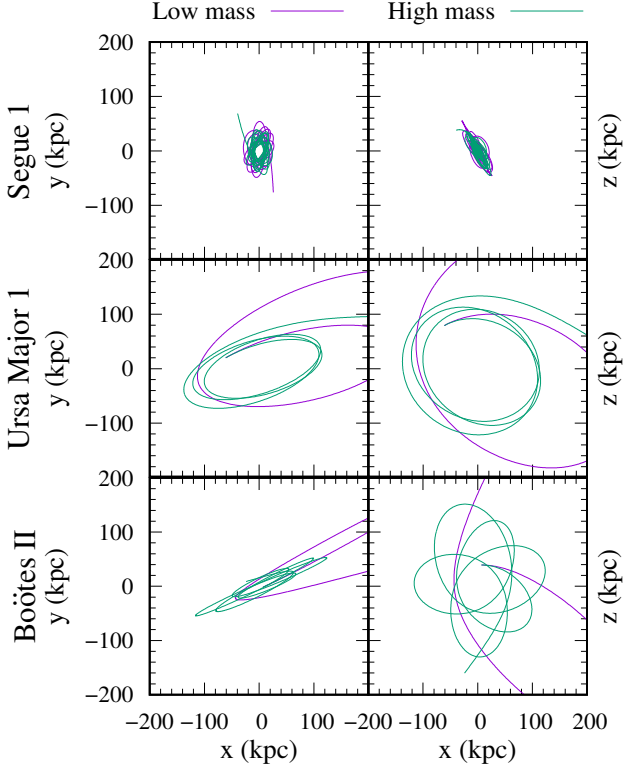}
    \caption{
    Three example UFD trajectories (Segue 1, UMa I, and Bo\"o II) for the lowest mass 
    ($M_{\rm h}=1.0\times10^{12}{\rm M}_{\odot}$; 
    purple) and highest mass ($M_{\rm h}=2.2\times10^{12}{\rm M}_{\odot}$; green) potentials consistent 
    with \citet{Watkins19}. The left figures are in the xy-plane and the right 
    figures are in the xz-plane. These trajectories do not account for uncertainty. The UFDs have varied 
    paths that are not strongly tied to the plane of the disc.
    }
    \label{egufd}
\end{figure}

\begin{figure}
	\includegraphics[width=\columnwidth]{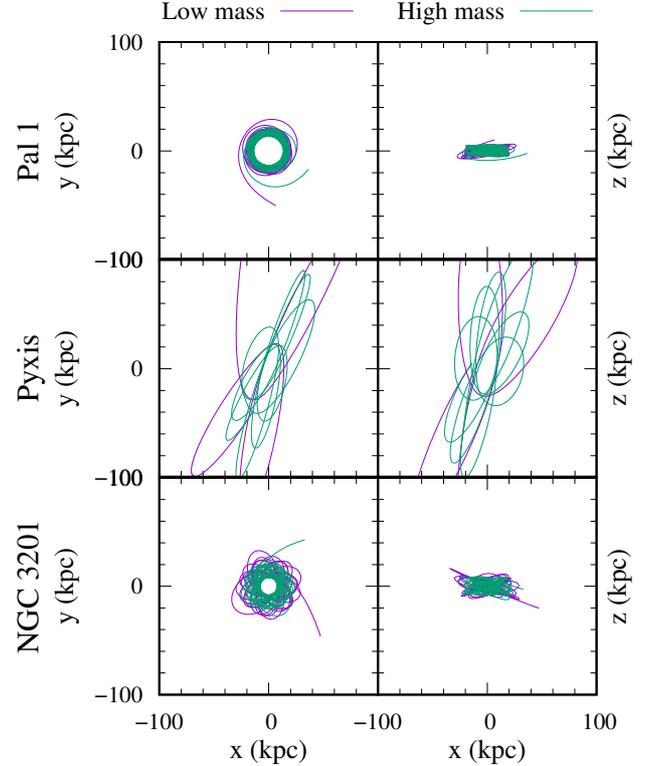}
    \caption{
    Three example GC trajectories (Pal 1, Pyxis, and NGC 3201) for the lowest mass 
    ($M_{\rm h}=1.0\times10^{12}{\rm M}_{\odot}$; 
    purple) and highest mass ($M_{\rm h}=2.2\times10^{12}{\rm M}_{\odot}$; green) potentials consistent 
    with \citet{Watkins19}. The left figures are in the xy-plane and the right 
    figures are in the xz-plane. These trajectories no not account for uncertainty. Most GCs follow the 
    example of Pal 1 and move in prograde around the disc, but clusters like Pyxis and NGC 3201 display 
    very different behaviour.
    }
    \label{eggc}
\end{figure}

To calculate the trajectory of a satellite, we first need to know its right ascension, declination, 
proper motions, distance, and radial velocity. We took these kinematic parameters and converted 
the positions and motions into Galactocentric coordinates\footnote{We have defined 
our coordinate system such that U velocities are positive towards the Galactic centre, V velocities are 
positive in the direction of Galactic rotation, and W velocities are positive towards the Galactic 
north pole. This convention is the same as in \citet{Gaia} and opposite that of \citet{Simon18}.} using
the Astropy software package for Python \citep{Astropy1, Astropy2}. We made the assumption that the
Sun is $d=8.3$ kpc from the Galactic centre \citep{Gillessen09} and 27 pc above the mid-plane \citep{Chen01}.
We have assumed a solar motion of $(U,V,W)_\odot=(11.1, 12.24, 7.25)$ km ${\rm s^{-1}}$ relative to the 
local standard of rest \citep{SBD10}, and that the MW has a circular velocity of 240 km ${\rm s^{-1}}$ 
at the distance of the Sun. We integrated the orbits of the satellites backwards for 13 Gyr using our 
own orbital integrator, which uses the second-order Leapfrog method, to determine the orbital histories 
of each satellite. Examples of UFD and GC trajectories are shown in Fig.~\ref{egufd} \& Fig.~\ref{eggc}, 
respectively.

In order to fully explore the available parameter-space of the satellites' orbits, we allowed the 
measurements for distance, radial velocity, and proper motions take five possible values: the given value, 
the value $\pm$ error/2, and the value $\pm$ error. 
As discussed above, because 101 of the GCs in \citet{Baumgardt18} did not have 
distance uncertainties we introduce an uncertainty of 10 per cent. While this is higher than the 
average distance error for the other GCs, we have leaned on the side of caution. 
There is no such problem with the UFDs in \citet{Simon18}.
We have ignored the error on the celestial coordinates because they are negligible. Exploring every 
combination of parameters, we have a total of $625$ possible trajectories for each UFD and GC per 
potential. Between the 10 different halo masses explored and the four potential models, this means 
that $25000$ trajectories were explored per satellite, for a total of $4275000$ orbits. 

We have treated every potential trajectory of a satellite for each model and halo mass as being 
equally likely. This treatment is unrealistic: there are two sets of orbits at the extremes of the uncertainty 
range but only one at the centre, which one might expect to be the most likely value for the parameter. 
Our results therefore may be biased towards more improbable, extreme orbits, while more conventional orbits may be 
under-represented. This means that a satellite that might otherwise be bound to the MW could have a 
specific combination of unlikely parameters come together to show a history of accretion. However, 
this also means that the absence of any signs of accretion despite the thorough exploration of 
parameter-space will be strong evidence that the satellite has been bound to the MW for an extended 
period of time. An important implication of this is that discussions about fractions of orbits are 
distinct from probabilities; a UFD for which 90 per cent of the orbits investigated are bound 
is not bound with a 90 per cent probability. Since those 10 per cent of clusters likely require extreme combinations 
of parameters at their utmost limits, the chances are very likely smaller.

\begin{figure*}
	\includegraphics[width=\textwidth]{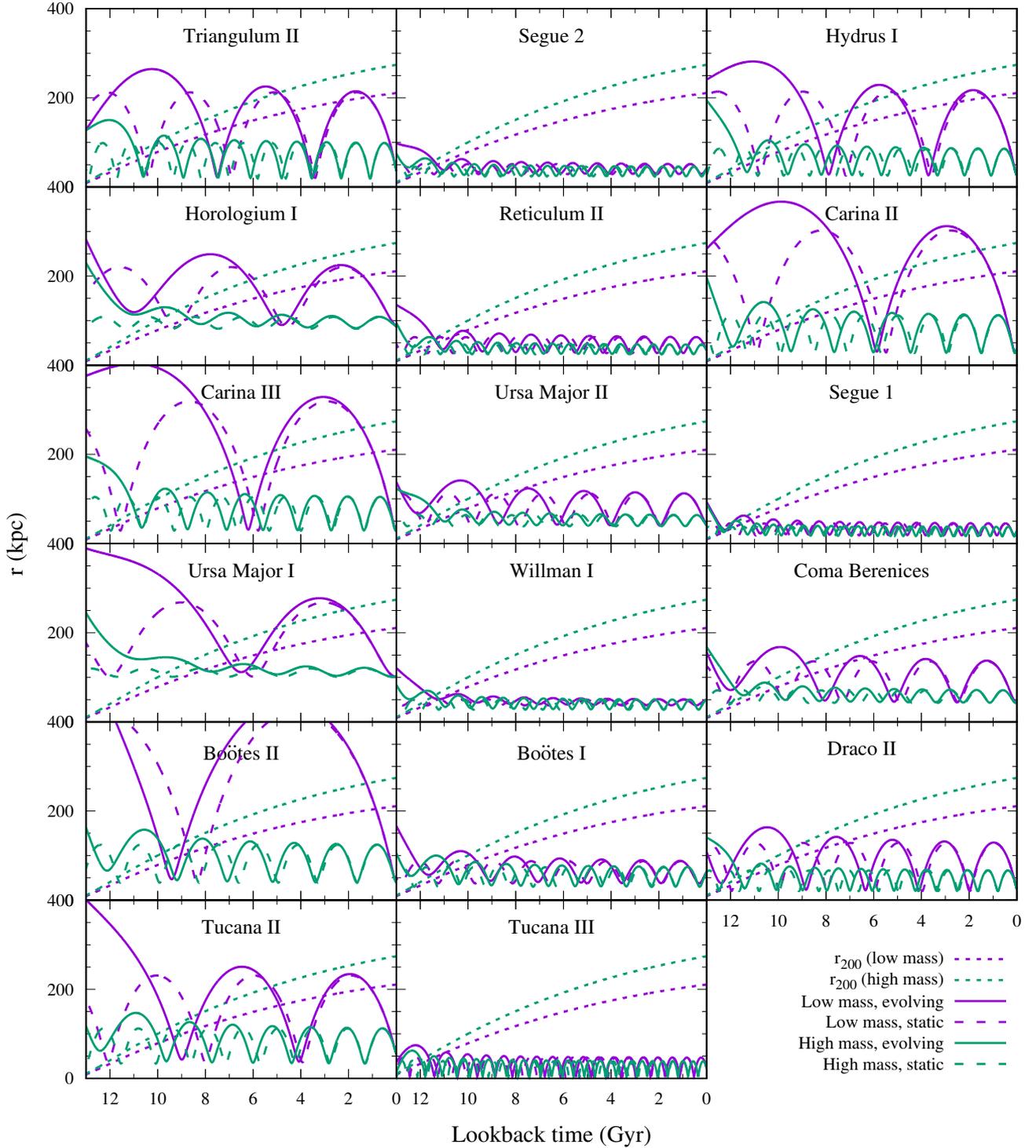}
    \caption{
    Galactocentric distance to the UFDs for the lowest mass ($M_{\rm h}=1.0\times10^{12}{\rm M}_{\odot}$; 
    purple) and highest mass ($M_{\rm h}=2.2\times10^{12}{\rm M}_{\odot}$; green) potentials consistent 
    with \citet{Watkins19}. The small difference is made up by $M_{\rm d}$. The solid and dashed lines 
    show M1, the evolving potential, and M2, the static potential, respectively. The dotted lines show the 
    virial radius of the halo for the high and low mass haloes throughout time. The first point at which 
    the orbits cross the virial radius is considered its infall time.
    }
    \label{orbits}
\end{figure*}

\section{Results}

After tracking each orbit, we calculate $f_b$, the fraction of possible orbits that are currently 
bound to the MW, for each satellite. We determine the present-day
pericentre, apocentre, eccentricity, and angle of the rotation vector
for each system by averaging the results over the most recent 2 Gyr of integration \citep{Gaia}, 
the results of which can be found in Tables~\ref{full_ufd} \& \ref{full_gc}.
We do this because the potential's asymmetry causes them to oscillate slightly between 
passes of the Galactic centre, occurring the most dramatically for closer GCs. Because 
these GCs also have short periods, 2 Gyr is sufficient to capture their full range of behaviours.
For orbits with periods $>2$ Gyr, we instead continued following 
their trajectories for one complete orbit.
We also calculate the infall time for each satellite by finding the earliest time 
at which it crosses the virial radius its parent dark matter halo. 
Satellites no measurable pericentre or apocentre are considered to be 
undergoing their first approach of the Galaxy.

Previous studies have claimed that a time-evolving potential can substantially change the trajectory 
of satellites around the Galaxy \citep{Zhang12, Haghi15, Miyoshi20}, but unless they are examined 
in the greater context of the existing parameter uncertainties, it is difficult to 
tell how reliable any conclusions are. Since one of the largest 
unknowns in the equation is the mass of the MW \citep[see e.g.][]{Yozin15}, this makes for a natural 
point of comparison. If we assume perfect information about the kinematic parameters 
and compare a static and evolving potential with the bounds over a realistic MW mass, 
we should be able to test how impactful it is. 
While some UFDs are known to be associated with the Magellanic Clouds, 
we presently neglect this point because even if they are not the `true' orbits, they can still 
show the impact of a changing potential on a satellite.
Fig.~\ref{orbits} plots the Galactocentric distance for each of the UFDs across the 13 Gyrs of backwards 
integration. The purple and green lines correspond to halo masses of 
$M_{\rm h}=1.0\times10^{12}{\rm M_{\odot}}$ and $M_{\rm h}=2.2\times10^{12}{\rm M_{\odot}}$, 
respectively, which, combined with the disc and bulge components used in M1 and M2 
(Table~\ref{Models}), closely matches the range of MW masses expected by \citet{Watkins19}. 
The solid lines represent M1, the evolving potential, while the dashed lines show M2, the static variant. 
The short-dashed lines indicate the time evolution of the virial radius, $r_{200}(z)$, for each model.

Comparing the range of trajectories for each UFD, it is apparent that most have 
substantially different orbits for the different values of $M_{\rm h}$. Since these are just the limits 
from \citet{Watkins19}, any similar orbit between these two extremes is plausible. 
The first pericentre for Tri II, Hyi I, Horo I, and Tuc II could reasonably be anywhere from $8-13$ 
Gyr ago and beyond, and Car II, Car III, UMa I, and Bo\"o II could be as recent as $6$ Gyr. 
By comparison, the difference between the evolving and static models is more subdued. 
The models start to noticeably diverge after $\approx 6$ Gyr, but this is mostly restricted to the 
low-mass cases where the UFDs are on very long orbits. These orbits can see a significant change 
in the timing of the first pericentre approach (e.g. UMa I), but this is certainly not guaranteed.
At the high-mass end, the difference is minimal until the final Gyrs, and even then only in some cases. 
The first pericentre approach happens at more or less the same time across the board. 
In roughly one third of the cases (Seg 2, Ret II, Seg 1, Wil I, Bo\"o I, \& Tuc III), 
the orbit is bound tightly enough that neither changing the mass nor evolving the potential 
greatly affects the trajectories. Because of their proximity and relatively stable orbits, 
most GCs also fall into this category, so we do not provide the same plots for the 154 Galactic 
GCs studied in this work.

While the evolution of the potential can have an effect on a satellite's trajectory, 
particularly when going back to the very early Universe, it is significantly less 
important than the present-day structure of that potential being tested. If we loosen 
our constraints and allow for uncertainty in the trajectories and for the model of the 
potential to vary in structure, not just halo mass, then we allow for an even wider range 
of possible trajectories, minimising the role of the potential evolution. 
Any attempt to draw conclusions from an orbit based upon the inclusion of a time-varying 
potential should therefore be treated with caution. This is not to say that a time-evolving potential 
is worthless and should be neglected in future studies, just that it is a relatively small 
problem compared to the existing issue of correctly modelling the mass distribution of the Galaxy.

\begin{figure*}
	\includegraphics[width=\textwidth]{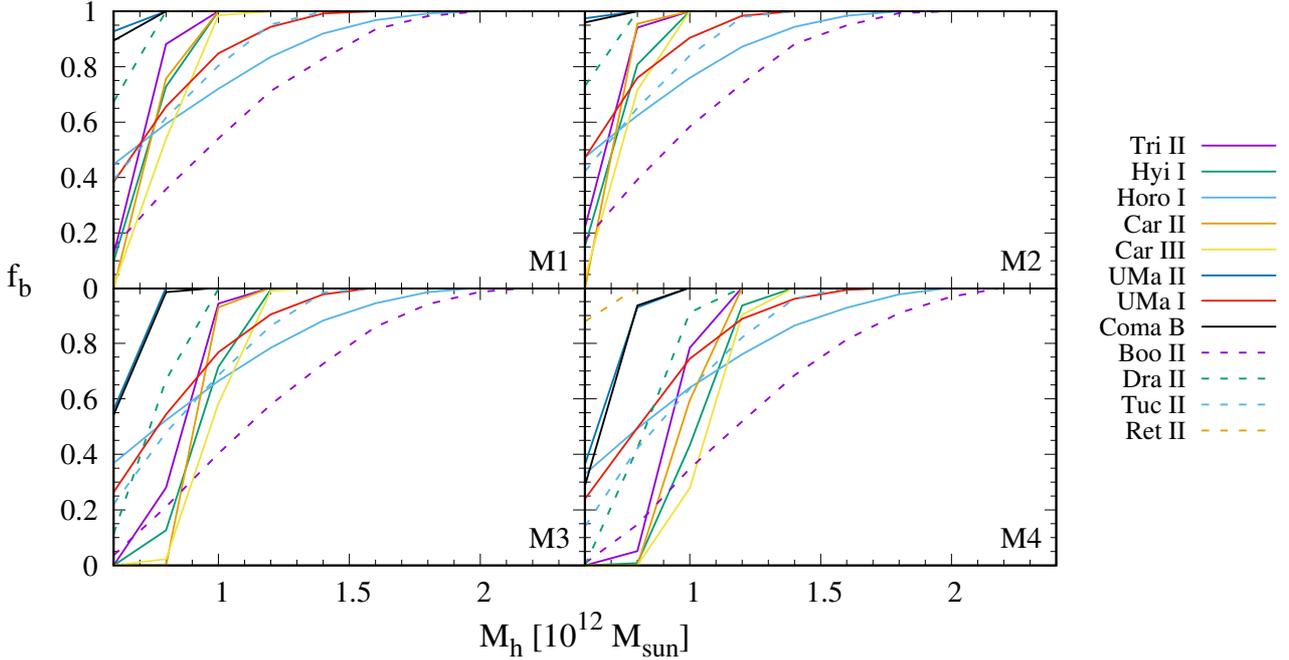}
	\vspace{-5mm}
    \caption{The fraction of investigated UFD trajectories that are bound to the MW as a function of the present-day 
    mass of the dark matter halo. UFDs which are always bound are not displayed. Top-left: model M1, an evolving 
    potential with a large disc and bulge component; top-right: M2, a static potential with large disc and bulge; 
    bottom-left: M3, an evolving potential with a lower mass disc and bulge; bottom-right: M4, an evolving halo-only 
    potential. Each model is described in Table~\ref{Models}.}
    \label{FA_UFD}
\end{figure*}

\begin{figure*}
	\includegraphics[width=\textwidth]{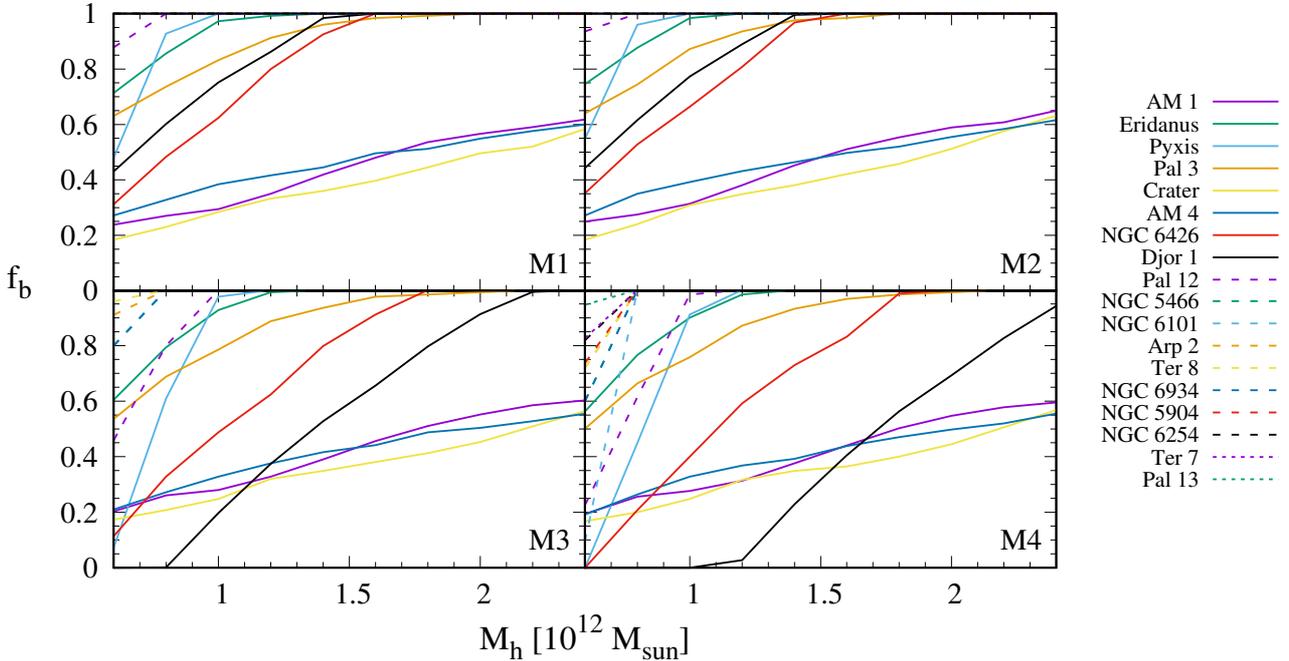}
	\vspace{-5mm}
    \caption{The fraction of investigated GC trajectories that are bound to the MW as a function of the present-day 
    mass of the dark matter halo. GCs which are always bound are not displayed. Top-left: model M1, an evolving 
    potential with a large disc and bulge component; top-right: M2, a static potential with large disc and bulge; 
    bottom-left: M3, an evolving potential with a lower mass disc and bulge; bottom-right: M4, an evolving halo-only 
    potential. Each model is described in Table~\ref{Models}.}
    \label{FA_GC}
\end{figure*}

\subsection{Fraction of bound orbits}

\subsubsection{Ultra-faint dwarf galaxies}

Fig.~\ref{FA_UFD} tracks $f_b$ for the UFDs as a function of halo mass using the four models 
described in Table.~\ref{Models}.
Only the 12 satellites with at least one unbound orbit are included; the other 5 UFDs always have $f_b=1$.
Comparing how $f_b$ changes in Fig.~\ref{FA_UFD}, M1 (an evolving potential)
and M2 (a static potential) are almost identical. This is unsurprising.
We showed in Fig.~\ref{orbits} that the difference between an evolving and a static 
potential is only really prominent in the earliest couple of Gyr, so for a UFD to be on first approach in 
one case but not the other would require the unusual circumstance where the UFD in the static case has 
its only other pericentre approach during this small window. As can be inferred from Fig.~\ref{orbits}, 
this is viable for very few of the UFDs.

The difference between M1 and M3, which has a smaller bulge and disc component, or M4, which 
consists of a dark matter halo only, are significantly more pronounced. 
Where Fig.~\ref{orbits} showed that the halo mass was more important than its evolution, Fig.~\ref{FA_UFD} 
shows that the same argument can be made about the way that $M_{\rm MW}$ is divided up into its components.
The most significant factor determining whether a UFD is accreting or bound is the initial value for $M_{\rm h}$. 
There are 11 UFDs in M1 that have at least one possible trajectory consistent with a first approach, with a 
12th introduced in M4. While this is a large subset of the initial population of 17, the number of accretion-capable 
UFDs drops dramatically as $M_{\rm h}$ increases. In M1, only five UFDs can accrete by 
$M_{\rm h}=1.0\times10^{12}{\rm M}_{\odot}$, and none by $M_{\rm h}=2.0\times10^{12}{\rm M}_{\odot}$.

The UFDs can be divided into two broad categories.
Eight UFDs see a very fast jump from $f_b\approx0$ to $f_b\approx1$ over one or two $M_{\rm h}$ intervals. 
These systems are usually always bound or always accreting, rarely in-between. These are Triangulum II, 
Hydrus I, Carina II, Carina III, Ursa Major II, Coma Berenices, and Draco II. The second category are 
the UFDs that also experience a noticeable change in $f_b$ between $M_{\rm h}$ values and models, but 
do so gradually. These include the four UFDs Horologium I, Bo\"otes II, Ursa Major I, and Tucana II.
The final UFD, Reticulum II, is not present for long enough to clearly fall into either category.
Looking at these satellites in greater detail \citep[Table~\ref{full_ufd}]{Simon18},
the key difference is in the uncertainties. Gradually increasing UFDs have at least one parameter with an 
unusually large uncertainty, whereas those that change quickly do not. This is particularly prominent
for the proper motions and radial velocities where large uncertainties tend to cause the velocity vector, 
and thus the energy of the satellite, to be overestimated \citep{Li20a}. This results in a wider range of possible 
trajectories, causing $f_b$ to change more gradually with $M_{\rm MW}$.

Some of these satellites are of particular interest. If we examine only the non-Magellanic UFDs, 
the five UFDs with a rapid change in $f_b$ are Tri II (purple solid),
Hyi I (green solid), UMa II (dark blue solid), Coma B (black solid), and Dra II (green dashed). 
While Coma B and Hyi I have very a
similar $f_b$, Tri II and Dra II are noticeably different. Each of these satellites are highly sensitive to 
$M_{\rm h}$, going from almost always accreting to almost always being bound over two mass intervals. 
This means that these satellites can be used to place limits on the mass of the Galaxy; if we can 
independently check whether these UFDs have recently accreted or not, we can constrain the MW mass to only 
values that can reproduce the UFD's history. \citet{Boylan-Kolchin13} used the velocity of the 
dwarf spheroidal galaxy Leo I to place constraints on the MW mass, so similar studies with UFDs should 
be possible (but see \citet{Erkal20} on how the absence of the LMC may alter the outcome).

The other UFD of particular note is Bo\"otes II. Of all the UFDs, Bo\"o II has consistently has 
the lowest $f_b$ over most values of $M_{\rm h}$ tested. Despite having significant uncertainties, 
Bo\"o II is the UFD most likely to be on its first approach of the Galaxy. 
Conversely, if Bo\"o II is not accreting, then it is unlikely 
that any other UFDs are doing so either. Looking into the individual orbits in M1, for haloes of 
$M_{\rm h}\leq1.2\times10^{12}{\rm M}_{\odot}$ there are multiple combinations of parameters within the more 
likely central regions of the uncertainties that put Bo\"o II on its first approach of the Galaxy. Beyond 
this point, the accreting orbits require more improbable circumstances, with 
unusually high distances ($d=42.0\pm2.0~{\rm kpc}$), unusually low proper motions in Right Ascension 
($\mu_\alpha {\rm cos}\delta=-2.517\pm0.325~{\rm mas~yr}^{-1}$), or both. 
Bo\"o II is therefore a natural target for studies trying to distinguish between accreted and bound UFDs.

\subsubsection{Globular clusters}

Fig.~\ref{FA_GC} tracks $f_b$ using the GCs as a function of halo mass using the four models 
described in Table.~\ref{Models}. Only 18 of the clusters are displayed; the rest are always bound to the MW.
M1 and M2 are even more similar to each other than for the UFDs. This happens because the GCs are usually far 
closer to the Galactic centre, so their tightly-bound orbits are harder to disrupt.
M3 and M4 are even more different, with 9 of the GCs with $f_b\neq1$ only appearing in these two models. 
As with the UFDs, the evolution of the potential is insignificant compared to $M_{\rm h}$. The 18 GCs that 
can be accreting drop to just four by $M_{\rm h}=2.4\times10^{12}{\rm M}_{\odot}$.

Dividing the GCs up like the UFDs, five GCs see a quick jump (Pyxis, NGC 5466, NGC 6101, NGC 6934, Pal 12),
four GCs are gradual (Eridanus, Pal 3, NGC 6426, Djor 1), six do not have enough information 
(Arp 2, Ter 8, NGC 5904, NGC 6254, Ter 7, Pal 13), and three clusters change extremely slowly (AM 1, AM 4, Crater).
Compared to the UFDs, 
the fraction of GCs potentially on first approach is much smaller, but almost all of those clusters 
have a much more gradual case, particularly outside of M4. This reflects the higher number of GCs 
with large uncertainties. 

The cluster Djor 1 stands out as changing drastically between models, 
varying between $f_b=0.03$ in M1 to $f_b=0.86$ in M4 for a $M_{\rm h}=1.2\times10^{12}{\rm M}_{\odot}$ halo, 
setting it apart from every other cluster. 
This happens because Djor 1, while having large uncertainties, is also moving very quickly relative to 
similar clusters \citep{Baumgardt18}. 
Additionally, its proximity to the Galactic centre makes it highly susceptible to the asymmetric potential 
of the disc component.

In addition to the two types of orbits seen in the UFDs, we also have the three unusual clusters AM 1, AM 4, 
and Crater. These GCs all go from about $f_b=0.2$ to $f_b=0.6$ over the 
halo mass range, regardless of the model tested. 
This gentle slope makes the clusters appear visually distinct from any of the other satellites. 
The reason these clusters take this shape is similar to the slowly-changing clusters: they have unusually 
large uncertainties, even moreso than their counterparts.
AM 1 and Crater are also very far away from the Galactic centre compared to other GCs at 123.3 and 145.0 kpc,
respectively, while AM 4 has a 3D velocity of $\approx 450 {\rm km s^{-1}}$, with uncertainties 
of $>100{\rm km s^{-1}}$ in each direction \citep{Baumgardt18}. 
The placement of these clusters in Figs.~\ref{FA_UFD} \& \ref{FA_GC} are a clear example of the bias 
introduced by the way we select trajectories. For satellites with sufficiently wide error 
bars, it becomes increasingly likely that at least some combinations of parameters will produce a trajectory 
on its first approach of the MW while others will be bound. 
As a result, these satellites will tend to say around the middle of the plot, never becoming fully bound 
or wholly accreted. This matches the behaviour of AM 1, AM 4, and Crater. 

The GCs cannot be as easily used to constrain $M_{\rm MW}$ as the UFDs. In M1 and M3, the more realistic models, 
Pyxis is the only GC that has the requisite rapid change. The other clusters with a quick change in M4 
(NGC 5466, NGC 6101, NGC 6934, Pal 12) are potentially interesting if they display any evidence of accretion, 
but because the halo-only M4 is not an incredibly realistic model, their behaviour may not be reproduced by 
lowering the halo mass beyond our lower limit. Pal 12 (purple dashed), for example, changes very quickly 
in M4 but does so far slower in M3, so similar behaviour could occur for the other clusters. Because of the 
large uncertainties, we cannot make conclusive statements about which clusters are likely to have been accreted. 

\begin{figure}
	\includegraphics[width=\columnwidth]{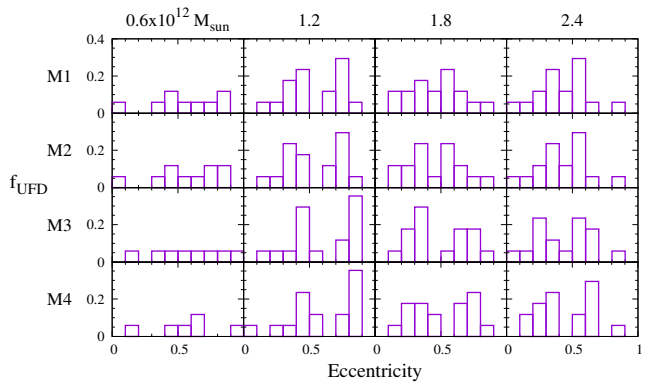}
    \caption{Histograms of the fraction of UFDs, $f_{\rm UFD}$, within each 0.1-width eccentricity bin. 
    Results are presented for four equally spaced halo masses, $M_{\rm h}$, for each model tested 
    (see Table~\ref{Models} for details). Unbound UFDs are not displayed, but are considered 
    when calculating $f_{\rm UFD}$. While the eccentricity distributions for M1 and M2 are almost identical, 
    M3 and M4 have noticeably less bound UFDs and the distribution shifts to favour higher eccentricity orbits.
    The halo mass, $M_{\rm h}$, causes the most significant change by binding more UFDs and lowering eccentricities 
    as $M_{\rm h}$ increases. The evolution of the potential is insignificant compared to how the potential is modelled.
    Although there are only a small number of UFDs, we see similar trends with the more populous GCs (Fig.~\ref{hist_gc}).
    }
    \label{hist_ufd}
\end{figure}

\subsection{Orbital eccentricities}

\begin{figure}
	\includegraphics[width=\columnwidth]{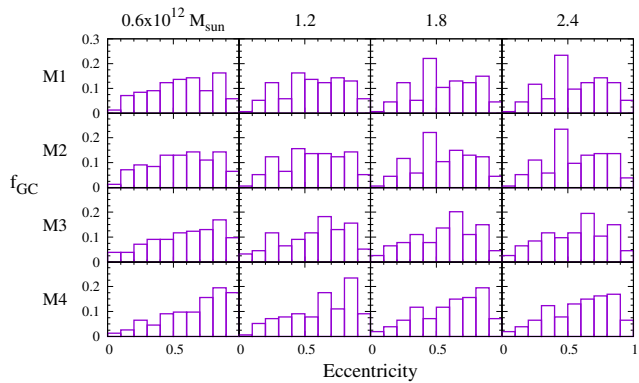}
    \caption{Histograms of the fraction of GCs, $f_{\rm GC}$, within each 0.1-width eccentricity bin. 
    Results are presented for four equally spaced halo masses, $M_{\rm h}$, for each of the four models tested 
    (Table~\ref{Models}). Unbound GCs are not displayed, but are considered 
    when calculating $f_{\rm GC}$. Like the UFDs in Fig.~\ref{hist_ufd}, eccentricity decreases as $M_{\rm h}$ rises, 
    increases in models with smaller     disc and bulge components, and is nearly unaffected by the evolution of the potential.
    }
    \label{hist_gc}
\end{figure}

To investigate the behaviour of individual orbits in more depth, we created a histogram of the eccentricities of each 
UFD (Fig.~\ref{hist_ufd}) and GC (Fig.~\ref{hist_gc}) for four equally-spaced values of $M_{\rm h}$ is each model. 
The fraction of UFDs or GCs across all eccentricity bins does not always sum to one; unbound satellites have no 
measurable eccentricity and thus are excluded by necessity. This is most obvious for the UFDs in Fig.~\ref{hist_ufd} where, 
due to the low number of total UFDs, one can clearly see the number of satellites in each bin.
Now that we are able to see the change in individual orbits instead of results averaged over all trajectories in 
Figs~\ref{FA_UFD} and \ref{FA_GC}, we can see a slight difference between M1 and M2 
for both kinds of satellites.
However, both models continue to follow the same trends, with only minor 
variations in each bin. By comparison, M3 and M4 see a much more significant change, with eccentricities shifting 
predominantly towards higher values, particularly in the case of the GCs. The biggest difference 
between these models is the effect that the disc has in introducing a non--spherically symmetric 
potential component. The disc pulls the satellites into a similar orbit, lowering their eccentricities.
This is why the clusters are affected more strongly; their proximity to the disc 
means that their orbits are significantly more affected by the asymmetry, causing a more severe eccentricity drop.

The biggest factor affecting the histogram is again the halo mass, $M_{\rm h}$. Higher $M_{\rm h}$ histograms 
have a larger proportion of satellites with eccentricities that are not very high or very low, 
favouring more central values. This is obscured somewhat by the changing number of accreting UFDs 
in the lower halo masses, but is clearer when looking at the GCs (Fig.~\ref{hist_gc}). 
The biggest influence on individual orbits are the different components of the potential, 
how they're constructed, and the masses used. By contrast, the evolution of that potential 
has a much smaller effect on the behaviour of the orbits. 
This is consistent with our previous findings that changing the depth of the potential through evolution 
is far less important than accurately modelling the present-day shape of the potential.
Until we have a more complete image of the mass distribution of the MW, modelling the  
evolution of the potential in detail is unlikely to significantly influence the outcome of satellite orbits.

\begin{figure*}
	\includegraphics[width=\textwidth]{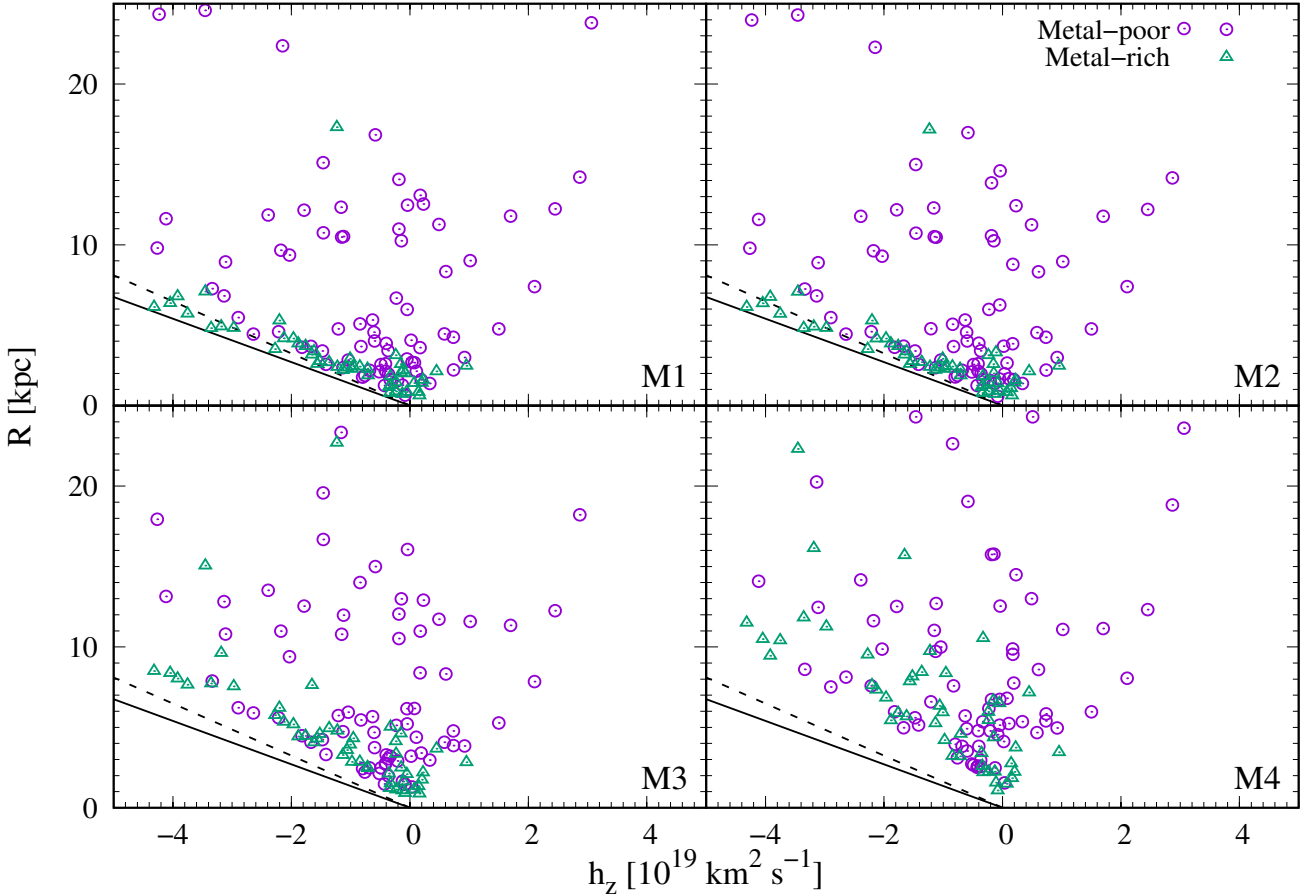}
	\vspace{-5mm}
    \caption{Specific angular momentum along the \textit{z}-axis, $h_z$, of the inner MW GCs compared with $R$, the 
    2D distance from the Galactic centre in the plane of the disc. $R$ is averaged over 2 Gyr. 
    Models are described in Table~\ref{Models}.
    We have used a halo mass of $M_{\rm h}=1.8\times10^{12}{\rm M}_{\odot}$. 
    A positive $h_z$ indicates the cluster is moving in retrograde to the disc. 
    A negative $h_z$ means prograde motion.
    The GCs have been divided into metal-poor (${\rm [Fe/H]}\leq-1.1$ dex; purple circles) and 
    metal-rich (${\rm [Fe/H]}>-1.1$ dex; green triangles) populations.
    The solid line and the dashed line show the $h_z$ of a satellite with a circular velocity of 240 and 200 ${\rm km~s^{-1}}$, 
    respectively.
    }
    \label{Metal}
\end{figure*}

\begin{figure}
	\includegraphics[width=\columnwidth]{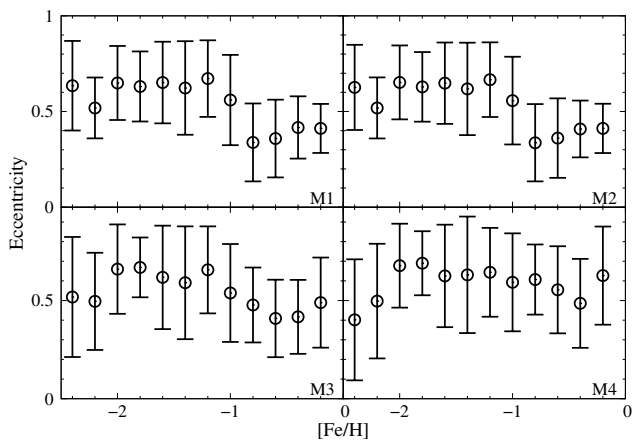}
	\vspace{-5mm}
    \caption{
    Eccentricity distribution of GCs divided into 0.2-width metallicity bins for a 
    $M_{\rm h}=1.8\times10^{12}{\rm M}_{\odot}$ halo. The circle and bars indicate the 
    mean value and standard deviation of the bin, respectively. The higher metallicity clusters show a clear 
    preference for lower eccentricity orbits. M1 and M2 look like two separate populations of $e\approx 0.65$ and 
    $e\approx 0.35$, while M3 has a more gradual downwards trend. M4 does not show a substantial change with 
    metallicity.
    The slight dip at the low-metallicity end is likely an artefact from the small number of clusters with 
    ${\rm [Fe/H]}<-2.1$.
    }
    \label{e_z}
\end{figure}

\begin{figure*}
	\includegraphics[width=\textwidth]{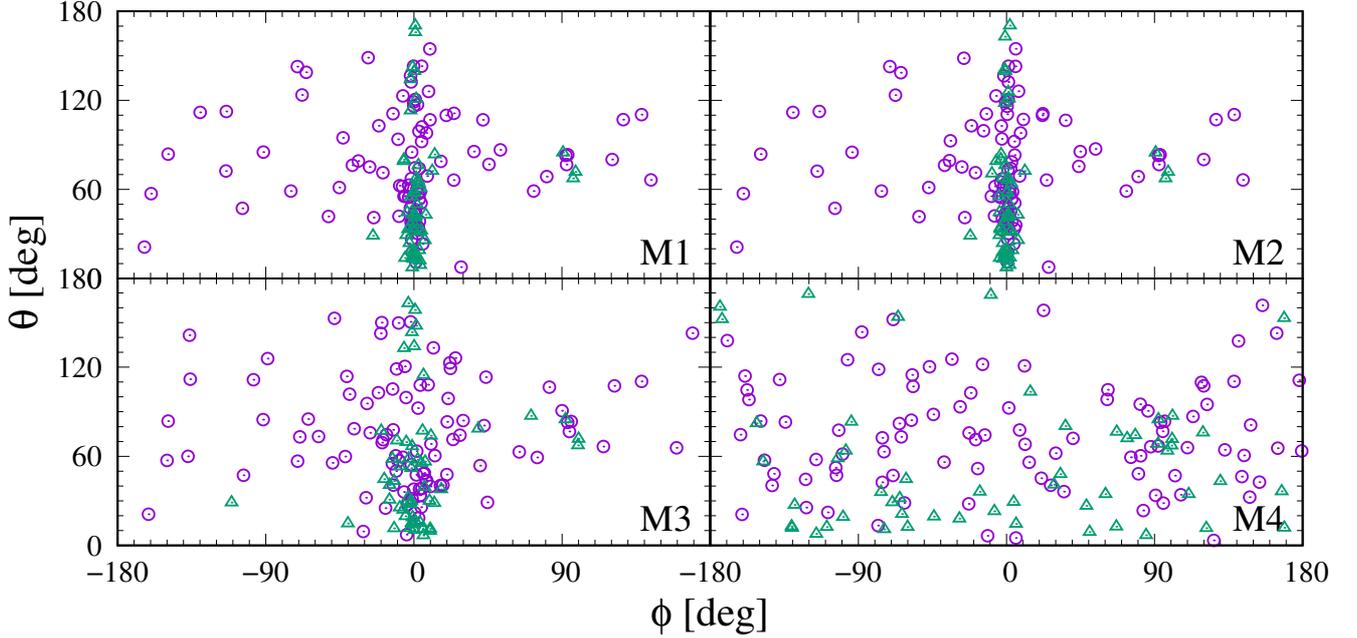}
	\vspace{-5mm}
    \caption{
    Angle of the GC rotation vector. $\theta$, the mean inclination, is measured from the Galactic north pole. 
    $\phi$ is the mean angle measured clockwise from the direction of the Galactic centre from the Sun to the 
    projection of the rotation vector onto the plane of the disc. Both values are averaged over 2 Gyr. 
    The asymmetric disc potential in M1-M3 causes $\phi$ of close satellites to approach zero as the orbital 
    plane moves around the $z$-axis. M4 is symmetric, so the rotation vector does not change. This means it 
    acts as a snapshot of the present-day values. While many metal-poor clusters do not have a strong connection 
    to the potential of the disc, almost all of the metal-rich clusters do.
    }
    \label{spin_gc}
\end{figure*}

\begin{figure*}
	\includegraphics[width=\textwidth]{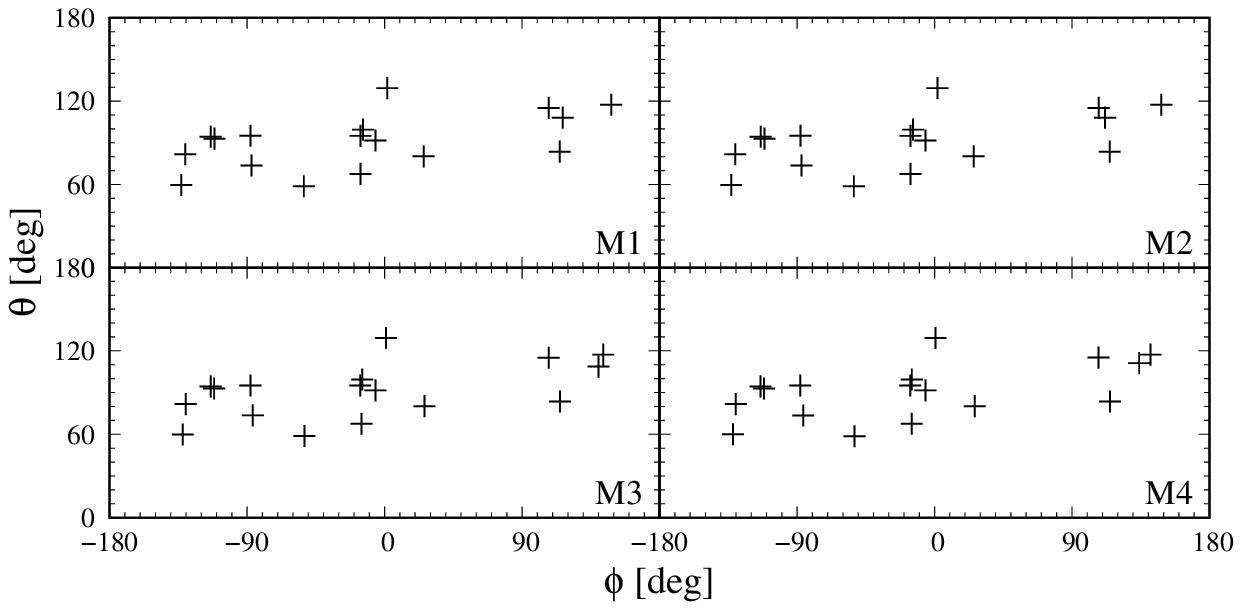}
	\vspace{-5mm}
    \caption{
    Angle of the UFD rotation vector. $\theta$, the mean inclination, is measured from the Galactic north pole. 
    $\phi$ is the mean angle measured clockwise from the direction of the Galactic centre from the Sun to the 
    projection of the rotation vector onto the plane of the disc. Both values are averaged over 2 Gyr. 
    While the asymmetric disc potential woud cause nearby satellites to cluster at $\phi=0^{\circ}$ as in 
    Fig.~\ref{spin_gc}, no UFDs are close enough for this effect to matter. Many of 
    UFDs are orbiting within $60^{\circ}<\theta<120^{\circ}$, nearly perpendicular to the plane of the disc.
    }
    \label{spin_ufd}
\end{figure*}

\subsection{Metal-rich and metal-poor cluster kinematics}

\subsubsection{Globular cluster motion in the disc}

While we did not find clear evidence that there were any GCs currently undergoing accretion, 
we can still explore the kinematic features of the metal-rich (${\rm [Fe/H]}>-1.1$; green triangles) and
metal-poor (${\rm [Fe/H]}\le-1.1$; purple circles) GC populations, where this threshold was chosen to 
be roughly halfway between the two metallicity peaks \citep[2010 edition]{Harris96}.
We measured the specific angular momentum projected along the $z$-axis, $h_z$, of each GC orbit and compared that with the 2D 
distance along the plane of the disc, $R$, dividing the results into metal-poor and metal-rich. 
Our results are shown in Fig.~\ref{Metal}. 
Four clusters with no metallicity measurements were excluded (Crater, FSR 1716, FSR 1735, and Mercer 5).
We used the model M1 with a halo mass of $M_{\rm h}=1.8\times10^{12}{\rm M}_{\odot}$. Changing $M_{\rm h}$ 
does not substantially change these results. The black solid line 
indicates what the $h_z$ of a satellite at that distance would be if travelling with a circular velocity of 
240 ${\rm km s^{-1}}$, and the black dashed line does the same for 200 ${\rm km s^{-1}}$. This approximately covers the range of the 
MW rotation curve. $R$ is averaged over 2 Gyr. Clusters beyond 25 kpc are not shown. Because of the 
chosen coordinate system, a negative $h_z$ indicates prograde motion while a positive value is retrograde.

The difference between the different metallicity populations is clear. In M1 and M2, the metal-rich clusters form a  
line that closely follows the black lines, which roughly indicating the motion of the disc. There is some variation
for clusters within about 5 kpc, but otherwise they almost universally move very closely with the disc. Moving 
to M3, we see more scatter and the gradient formed is steeper, and finally M4 has a wide spread of 
$R$. By contrast, the metal-poor clusters do have a noticeable number of clusters following the same 
line, but many others are scattered throughout the graph on both the prograde and retrograde side. These clusters 
do not show the same movement between models, which indicates that they do not have the same strong connection 
to the disc. This also happens independently of how close a satellite is to the centre of the disc.

If a GC were to be formed \textit{in-situ}, we would expect it to closely follow the kinematics of the disc. 
This is clearly seen in the metal-rich clusters in M1, which means that these high metallicity clusters have 
most likely formed within the Galactic thin disc. This may imply that the metal-poor population has
a significantly different origin. Because the metal-poor GCs that are not lying on the black lines are fairly 
evenly distributed between prograde and retrograde and do not show obvious trends that might indicate a common 
origin, it is plausible that these clusters accreted onto the Galaxy in the past.

\subsubsection{Eccentricity-metallicity relation}

While we have used a sharp cut-off to divide the metal-rich and metal-poor clusters, the bimodality 
is best represented by a pair of overlapping Gaussians and the [Fe/H] = -1.1 threshold is not  
physically-motivated beyond being roughly halfway between their peaks. It is unclear how 
well this division separates the populations. To investigate whether cross-over in the populations
is muddling the results, we searched for a connection between the eccentricity and metallicity 
for each model.
Fig.~\ref{e_z} shows the eccentricity and metallicity for each model, divided into 0.2-dex bins. 
The circle indicates the mean value in the bin and the bars the standard deviation. We have 
assumed a halo mass of $M_{\rm h}=1.8\times10^{12}{\rm M}_{\odot}$. As before, changing $M_{\rm h}$ 
does not substantially change these plots.

In M1 and M2, the metal-rich and metal-poor clusters behave very differently. The metal-poor clusters 
predominantly have eccentricities of $e\approx 0.65$, while the metal-rich clusters are closer to $e \approx 0.35$.
Interestingly, both populations are fairly unchanging, until we see a slight drop at -1 and a huge drop at -1.2. 
This clear division indicates that the clusters can be well described as being separated at about -1.0 or -1.1. 
While M3 has a gentler drop that levels out sooner and M4 shows no such drop, these two models are unlikely 
to be realistic because of the large disconnect from the disc in Fig.~\ref{Metal} for M3 and the fact that M4 
is a dark matter halo only. However, the fact that this difference is not reproduced in the disc-less M4 
indicates that this eccentricity distribution is not an inherent feature of the clusters, but a result 
of their place in the Galactic potential.
As we saw above in Fig.~\ref{hist_ufd} \& \ref{hist_gc}, the potential of the disc 
lowers the eccentricities of nearby orbits. The metal-rich clusters are predominantly located towards 
the centre of the Galaxy (Fig.~\ref{Metal}), where they are much more likely to have their 
eccentricities suppressed. There is a slight dip at the low metallicity-end, but this is likely due to the 
very small number of clusters within those bins.

\subsection{Angle of rotation}

Another property of the satellites we can examine is the orientation of the averaged rotation vector,
the momentary orbital pole about which the satellite rotates clockwise.
This is defined by $\theta$, the mean inclination from the Galactic north pole, and $\phi$, 
the mean angle clockwise around the disc from the direction of the Galactic centre from the 
Sun's current position to the projection of the orbital rotation vector onto the plane of the disc, 
both averaged over 2 Gyr. Figs.~\ref{spin_gc} \& \ref{spin_ufd} show the mean angle of the rotation vector 
for the GCs and UFDs, respectively, with the GCs split into metal-rich and metal-poor populations. 
We average the rotation vector because the asymmetric potential of the disc will cause satellites 
that approach close enough to vary this vector with time. The additional disc component of the potential
will cause the satellite to move between its furthest points above and below the disc more rapidly,
causing the orbital plane to precess around the $z$-axis. $\phi$ can then take any value, so when averaged will 
tend towards $0^{\circ}$. We can confirm this by examining M4 in Fig.~\ref{spin_gc}, 
which has a dark matter halo--only potential.
We do not see the clustering towards a mean $\phi=0^{\circ}$ as we do in the 
other models because the rotation vector does not change with time. M4 therefore 
effectively acts as a snapshot of the present-day values.

\subsubsection{Metal-rich and metal-poor clusters}

The metal-rich (green triangles) and metal-poor (purple circles) satellites show very different behaviour. 
The metal-rich GCs cluster at $\phi=0^{\circ}$, so the disc's potential is strong enough to 
control the orbits. 47 of the 54 metal-rich clusters move in prograde.
The metal-poor GCs have a diverse range of orbits scattered across all $\phi$. While some lie upon 
$\phi=0^{\circ}$, a significant population does not. 48 of the 96 metal-poor clusters have orbital planes 
substantially offset from the plane of the disc ($60^{\circ} \leq \theta \leq 120^{\circ}$), a noticeably 
larger fraction than the 14 metal-rich clusters with similar orbits.

While most metal-rich clusters have $\theta \leq 30^{\circ}$, some move along substantially different orbits.
Seven metal-rich clusters have $\phi \approx 0^{\circ}$ but move in retrograde. These clusters are 
Ter 2 ($\theta=134^{\circ}, \phi=-2^{\circ}, {\rm [Fe/H]}=-0.69$), 
Lil 1 ($\theta=166^{\circ}, \phi=1^{\circ}, {\rm [Fe/H]}=-0.33$), 
NGC 6380 ($\theta=121^{\circ}, \phi=1^{\circ}, {\rm [Fe/H]}=-0.75$), 
NGC 6388 ($\theta=140^{\circ}, \phi=0^{\circ}, {\rm [Fe/H]}=-0.77$), 
NGC 6440 ($\theta=113^{\circ}, \phi=-2^{\circ}, {\rm [Fe/H]}=-0.36$), 
Ter 6 ($\theta=142^{\circ}, \phi=-1^{\circ}, {\rm [Fe/H]}=-0.56$), and 
2MASS-GC02 ($\theta=170^{\circ}, \phi=1^{\circ}, {\rm [Fe/H]}=-1.08$).

Of these clusters, NGC 6388, Ter 6, and 2MASS-GC02 are retrograde in each of their explored orbits, 
and NGC 6380, Ter 2, and Lil 1 are retrograde outside of extreme combinations of uncertainties, so 
these 6 clusters are most likely retrograde.
NGC 6440 is by default retrograde but shifts to a prograde orbit at larger initial distances. 
A better understanding of the distance uncertainty for this cluster will be required to work out 
the direction of its motion.

In addition to the retrograde clusters, there are four metal-rich GCs with unusually large $\phi$:
Whiting 1 ($\theta=72^{\circ}, \phi=98^{\circ}, {\rm [Fe/H]}=-0.70$),
E3 ($\theta=29^{\circ}, \phi=-25^{\circ}, {\rm [Fe/H]}=-0.83$),
Ter 7 ($\theta=85^{\circ}, \phi=91^{\circ}, {\rm [Fe/H]}=-0.32$), and
Pal 12 ($\theta=67^{\circ}, \phi=97^{\circ}, {\rm [Fe/H]}=-0.85$).

While 2MASS-GC02's metallicity is close enough to the ${\rm [Fe/H]}=-1.1$ cutoff that it could be 
misidentified, the other clusters are clearly metal-rich. These GCs are unlikely to have been formed 
within the disc and would require the accretion of a different kind of satellite to the kind resulting 
in metal-poor GCs, which make up the large majority of potentially accreting clusters. 
The Galaxy is thought to have undergone a merger with a massive, Magellanic Cloud--like dwarf galaxy 
with slight retrograde motion \citep{Belokurov18, Haywood18, Koppelman18, Helmi18}. Such a dwarf 
would likely be large enough to form such metal-rich GCs, making this a plausible scenario for the
origin for these clusters.

\subsubsection{Ultra-faint dwarf galaxies}

Unlike the clusters, the UFDs in Fig.~\ref{spin_ufd} show little difference between M1--3 and M4. 
Almost all of the UFDs 
have orbits too large to be significantly altered by the asymmetry of the disc potential, so they remain 
within the same orbital plane. This prevents the clustering at $\phi=0^{\circ}$ seen in the GCs. 
Almost all of the UFDs are found within $60^{\circ} \leq \theta \leq 120^{\circ}$, close to perpendicular 
to the plane of the disc. This is similar to observations of the MW classical satellites 
\citep{Metz08, Libeskind09, Deason11, Shao16}. While planes of satellites are common in other galaxies, 
both observations and simulations generally find them aligned with their host's disc
\citep[see e.g.][]{Brainerd05, Yang06, Agustsson10, Nierenberg12, Tully15, Welker18, Brainerd19}.
However, \citet{Cautun15} showed that planes of satellites have very diverse properties and that 
the Galaxy is just one possible outcome of this population, so cannot alone be used to test cosmological models.

While the orientation itself and the flatness of the satellite plane may be temporary and just a 
chance occurrence \citep{Shao19}, what is interesting is just how many satellites lie within this plane. 
\citet{Shao19} showed that 8 of the 11 classical dwarfs have likely shared the same orbital plane for 
billions of years. Given the similarities in orbits, the same is likely true for the UFDs.
This could make them useful for studying the structure of the dark matter halo. \citet{Shao20} argued 
that the MW halo was twisted such that the outer halo was perpendicular to the inner halo by studying 
MW analogues in simulations with a similar arrangement of classical satellites to the Galaxy. 
It would be interesting how the inclusion UFDs could help probe the halo in greater detail.

\section{Discussion and summary}

In this paper we have investigated how the evolution of the Galactic potential can influence the orbits 
of UFDs and GCs. We found that while an evolving potential can lead to noticeable differences in UFD orbits
when tracked backward for 6 Gyr or more, the differences are small compared to the existing uncertainties in 
orbital parameters. While one should account for the evolution where possible, one must consider 
this in the context of the full range of uncertainties. Because the relative impact is minor, we can justify using 
simple approximations to model the MAH of the Galaxy (e.g. COMMAH, \citealt{COMMAHc}).
Even if these models do not well represent the true MAH of the MW, they are almost certainly
closer than a static potential.

Our exploration of the 17 UFDs in \citet{Simon18} showed that, for model M1, 11 have at least one orbit 
consistent with a first approach scenario. Within the MW mass range predicted 
by \citet{Watkins19}, this drops to four: Bo\"otes II, Horologium I, Tucana II, and Ursa Major I.
By $M_{\rm h} = 1.4\times10^{12} {\rm M_{\odot}}$ ($M_{\rm MW} \approx  1.5\times10^{12} {\rm M_{\odot}}$), 
all accreting trajectories require unlikely parameter combinations. Bo\"otes II is the most likely 
candidate for being on first passage. Of the 154 Galactic GCs in \citet{Baumgardt18}, 9 have at at least 
one first approach trajectory in M1. Because of the uncertainties on these satellites, it is much harder 
to determine which, if any, are likely candidates.
The non-Magellanic UFDs Tri II, Hyi 1, Coma B, and Dra II and the GC Pyxis all swap from mostly bound 
to mostly accreting over a small mass range. If their accretion histories were independently verified, 
they could be used to place constraints on the MW mass.

We find the metal-rich and metal-poor GC sub-populations to be kinematically distinct. The metal-rich 
GCs behave consistently, with predominately prograde motion with respect to the disc, a clear relation 
between $R$ and $h_z$ (Fig.~\ref{Metal}), low eccentricities ($e\approx 0.35$, Fig.~\ref{e_z}),
and rotation vectors indicative of motion defined by the disc (Fig.~\ref{spin_gc}). This shows a 
fundamental link between these metal-rich GCs and the thin disc of the MW. By contrast, the metal-poor 
GCs show a diverse range of behaviours. They have higher eccentricities ($e\approx 0.65$), do not show the 
same connection to disc, and 31 of the 96 clusters are moving in retrograde. Half of the GCs are 
found within $60^{\circ} \leq \theta \leq 120^{\circ}$, which is where the UFDs are found
(Fig.~\ref{spin_ufd}). These features are consistent with a history of accretion.

This view of metal-poor GCs being accreted and metal-rich GCs forming \textit{in-situ}
presents a problem: the stars in the thin disc have higher metallicities than these GCs 
\citep[e.g.][]{Haywood02, Taylor05, Marsakov11, Schlesinger14, Hayden17}. If the GCs were formed from 
the same material, they should have similar metallicities. One possible solution is that the metal-rich 
gas in the MW was mixed with another source, such as an accreting metal-poor dwarf galaxy, which the GCs
were then formed from. This scenario was studied by \citet{BC02}, who found that this process
would form clusters of approximately [Fe/H] = -0.58, and that GCs formed by the dwarf before the merger 
would average [Fe/H] = -1.45. This is remarkably close to the -0.6 and -1.6 peaks of the MW clusters and 
may indicate that the metal-rich and metal-poor GCs shared an origin in the same accreting dwarfs. The 
gas would experience pressure where the clusters would not and be pulled into the rough plane of the disc, 
explaining why most metal-rich GCs orbit with a non-zero inclination (Fig.~\ref{spin_ufd}).

If there is a link between the disc and the metal-rich GCs, what can we say about the halo stars?
\citet{Carollo10} derived the orbital eccentricity for Galactic halo stars in various metallicity 
ranges. They found that stars with $-1.0$ < [Fe/H] < $-0.5$ peaked at $e \approx $ 0.2 and 
$e \approx $ 0.3 for ${\rm |z|} = 1-2$ and $2-4$ kpc, respectively. 
This matches the observed 
eccentricities for metal-rich GCs, suggesting that the metal-rich halo stars may have a common ancestry 
with the metal-rich GCs.
For halo stars with $-1.5$ < [Fe/H] < $-1.0$ and ${\rm |z|} = 1-2$ there was a small peak at $e \approx $ 0.35, 
but otherwise most of the stars were found at high eccentricities. While the peak points to some 
of these coming from mixed gas, most metal-poor halo stars follow the lead of the metal-poor GCs, 
suggesting that they are the stellar remnants of the accreted dwarfs.

There are 11 metal-rich clusters that do not fit neatly into this picture. Ter 2, Lil 1, NGC 6380, 
NGC 6388, NGC 6440, Ter 6, and 2MASS-GC02 move retrograde to the disc, while Whiting 1, E3, Ter 7, 
and Pal 12 do not have the $\phi=0^{\circ}$ expected of something created by the disc. While 2MASS-GC02 
has a metallicity close to the cutoff, the other GCs are clearly metal-rich. This may indicate that 
these clusters had a different kind of origin to the other GCs, forming within a more metal-rich, 
and thus due to the metallicity-mass relation of dwarf galaxies more massive, environment prior to accretion.
This would require in a significant retrograde merger occurring in the past, something supported 
by existing studies \citep[e.g. ][]{Belokurov18, Haywood18, Koppelman18, Helmi18}.
Since a surprisingly large number of metal-rich GCs have retrograde orbits, the accretion of retrograde 
cluster hosts may have a significant impact on Galaxy evolution.

\section*{Acknowledgements}
We thank the anonymous referee for their helpful comments and suggestions on how to improve this paper.

This research made use of Astropy,\footnote{http://www.astropy.org} a community-developed core Python package for Astronomy \citep{Astropy1, Astropy2}. 

ADL acknowledges financial support from the Australian Research Council through
their Future Fellowship scheme (project number FT160100250).








\appendix

\section{Data Availability Statement}

The data underlying this article will be shared on reasonable request to the corresponding author.

\section{Kinematic catalogue}

\setcounter{table}{0}
\renewcommand{\thetable}{A\arabic{table}}

\begin{table*}
	\centering
	\caption{Kinematic data for UFDs in M1 for the upper and lower mass bounds in \citet{Watkins19}.
	Uncertainties are 
	determined from the most extreme examples of orbits that are still bound. The mass of the bulge is 
	small and so has been neglected.}
	\label{full_ufd}
	\begin{tabular*}{\textwidth}{l @{\extracolsep{\fill}} ccllll}
    	\hline
        UFD & $M_{\rm h}+M_{\rm d} (\times10^{12} {\rm M_\odot})$  & $f_b$ & pericentre (kpc) & apocentre (kpc) & eccentricity & $t_{\rm infall}$ (Gyr)\\
		\hline
 Triangulum II & 1.1 & 1.00 & $ 20.8 _{ -4.0}^{+ 5.9 }$ &$ 214.8 _{ -70.6}^{+ 190.6 }$ &$ 0.82 _{ -0.03}^{+ 0.05 }$ &$ 8.0 _{ -2.3}^{+ 4.2}$\\
& 2.3 & 1.00 & $ 19.1 _{ -3.5}^{+ 4.1 }$ &$ 99.3 _{ -18.0}^{+ 29.0 }$ &$ 0.68 _{ -0.01}^{+ 0.02 }$ &$ 10.9 _{ -0.9}^{+ 1.5}$\\
 Segue 2 & 1.1 & 1.00 & $ 28.9 _{ -8.0}^{+ 7.6 }$ &$ 51.8 _{ -7.5}^{+ 18.7 }$ &$ 0.28 _{ -0.02}^{+ 0.08 }$ &$ 11.3 _{ -1.1}^{+ 0.5}$\\
& 2.3 & 1.00 & $ 23.8 _{ -6.6}^{+ 8.4 }$ &$ 45.9 _{ -3.8}^{+ 6.9 }$ &$ 0.32 _{ -0.08}^{+ 0.10 }$ &$ 11.5 _{ -0.2}^{+ 0.8}$\\
 Hydrus I & 1.1 & 1.00 & $ 26.3 _{ -1.1}^{+ 13.9 }$ &$ 217.4 _{ -95.0}^{+ 314.7 }$ &$ 0.78 _{ -0.13}^{+ 0.09 }$ &$ 8.4 _{ -7.6}^{+ 3.4}$\\
& 2.3 & 1.00 & $ 25.6 _{ -0.7}^{+ 0.7 }$ &$ 86.3 _{ -22.9}^{+ 35.4 }$ &$ 0.54 _{ -0.11}^{+ 0.10 }$ &$ 11.5 _{ -1.4}^{+ 0.5}$\\
 Horologium I & 1.1 & 0.72 & $ 90.1 _{ -44.4}^{+ 61.3 }$ &$ 224.8 _{ -145.4}^{+ 262.2 }$ &$ 0.43 _{ -0.30}^{+ 0.17 }$ &$ 5.7 _{ -5.1}^{+ 5.0}$\\
& 2.3 & 1.00 & $ 83.1 _{ -49.8}^{+ 26.1 }$ &$ 108.6 _{ -31.5}^{+ 309.4 }$ &$ 0.13 _{ -0.04}^{+ 0.45 }$ &$ 9.2 _{ -3.7}^{+ 2.3}$\\
 Reticulum II & 1.1 & 1.00 & $ 27.0 _{ -7.4}^{+ 5.7 }$ &$ 63.5 _{ -20.5}^{+ 49.8 }$ &$ 0.40 _{ -0.04}^{+ 0.15 }$ &$ 11.3 _{ -1.9}^{+ 0.7}$\\
& 2.3 & 1.00 & $ 24.7 _{ -7.5}^{+ 6.6 }$ &$ 46.5 _{ -9.4}^{+ 18.9 }$ &$ 0.31 _{ -0.01}^{+ 0.06 }$ &$ 11.4 _{ -0.5}^{+ 1.0}$\\
 Carina II & 1.1 & 1.00 & $ 28.4 _{ -1.9}^{+ 1.7 }$ &$ 312.1 _{ -83.8}^{+ 163.7 }$ &$ 0.83 _{ -0.04}^{+ 0.05 }$ &$ 6.5 _{ -0.7}^{+ 5.2}$\\
& 2.3 & 1.00 & $ 27.5 _{ -2.0}^{+ 1.9 }$ &$ 112.5 _{ -13.4}^{+ 18.5 }$ &$ 0.61 _{ -0.02}^{+ 0.03 }$ &$ 11.9 _{ -2.3}^{+ 0.1}$\\
 Carina III & 1.1 & 0.99 & $ 28.8 _{ -0.7}^{+ 0.7 }$ &$ 329.0 _{ -156.6}^{+ 541.7 }$ &$ 0.84 _{ -0.12}^{+ 0.10 }$ &$ 6.8 _{ -6.0}^{+ 4.9}$\\
& 2.3 & 1.00 & $ 28.9 _{ -0.8}^{+ 0.7 }$ &$ 104.7 _{ -27.1}^{+ 43.1 }$ &$ 0.57 _{ -0.10}^{+ 0.10 }$ &$ 11.0 _{ -1.4}^{+ 1.0}$\\
 Ursa Major II & 1.1 & 1.00 & $ 40.0 _{ -3.2}^{+ 3.2 }$ &$ 112.6 _{ -37.9}^{+ 80.1 }$ &$ 0.48 _{ -0.14}^{+ 0.16 }$ &$ 9.3 _{ -1.9}^{+ 2.0}$\\
& 2.3 & 1.00 & $ 38.9 _{ -4.2}^{+ 3.0 }$ &$ 64.0 _{ -14.2}^{+ 23.5 }$ &$ 0.24 _{ -0.07}^{+ 0.11 }$ &$ 11.3 _{ -1.1}^{+ 0.4}$\\
 Segue 1 & 1.1 & 1.00 & $ 18.1 _{ -6.9}^{+ 6.0 }$ &$ 45.4 _{ -10.5}^{+ 26.3 }$ &$ 0.43 _{ -0.02}^{+ 0.09 }$ &$ 11.1 _{ -0.7}^{+ 1.3}$\\
& 2.3 & 1.00 & $ 16.2 _{ -6.2}^{+ 6.5 }$ &$ 37.5 _{ -5.7}^{+ 11.6 }$ &$ 0.40 _{ -0.05}^{+ 0.13 }$ &$ 12.4 _{ -0.9}^{+ 0.3}$\\
 Ursa Major I & 1.1 & 0.85 & $ 101.8 _{ -11.7}^{+ 6.1 }$ &$ 277.6 _{ -177.8}^{+ 489.0 }$ &$ 0.46 _{ -0.42}^{+ 0.31 }$ &$ 7.0 _{ -6.3}^{+ 1.4}$\\
& 2.3 & 1.00 & $ 101.3 _{ -43.7}^{+ 6.6 }$ &$ 121.5 _{ -24.0}^{+ 238.5 }$ &$ 0.09 _{ -0.06}^{+ 0.45 }$ &$ 8.7 _{ -3.4}^{+ 1.8}$\\
 Willman 1 & 1.1 & 1.00 & $ 37.3 _{ -18.8}^{+ 22.5 }$ &$ 50.7 _{ -10.4}^{+ 77.8 }$ &$ 0.15 _{ -0.11}^{+ 0.22 }$ &$ 11.2 _{ -2.3}^{+ 0.7}$\\
& 2.3 & 1.00 & $ 27.6 _{ -12.5}^{+ 32.0 }$ &$ 50.3 _{ -10.1}^{+ 20.6 }$ &$ 0.29 _{ -0.25}^{+ 0.16 }$ &$ 11.4 _{ -0.8}^{+ 1.1}$\\
 Coma Berenices & 1.1 & 1.00 & $ 42.9 _{ -2.1}^{+ 2.1 }$ &$ 137.6 _{ -53.5}^{+ 133.5 }$ &$ 0.52 _{ -0.18}^{+ 0.19 }$ &$ 8.6 _{ -2.8}^{+ 2.4}$\\
& 2.3 & 1.00 & $ 42.9 _{ -2.9}^{+ 2.3 }$ &$ 71.5 _{ -19.2}^{+ 31.8 }$ &$ 0.25 _{ -0.12}^{+ 0.14 }$ &$ 10.5 _{ -0.6}^{+ 1.1}$\\
 Bo\"otes II & 1.1 & 0.54 & $ 46.7 _{ -11.1}^{+ 37.2 }$ &$ 446.7 _{ -351.0}^{+ 104.9 }$ &$ 0.81 _{ -0.36}^{+ 0.01 }$ &$ 9.7 _{ -9.2}^{+ 1.5}$\\
& 2.3 & 1.00 & $ 39.5 _{ -5.1}^{+ 5.3 }$ &$ 124.6 _{ -66.9}^{+ 314.1 }$ &$ 0.52 _{ -0.27}^{+ 0.30 }$ &$ 9.7 _{ -4.0}^{+ 2.0}$\\
 Bo\"otes I & 1.1 & 1.00 & $ 37.6 _{ -8.2}^{+ 8.4 }$ &$ 88.2 _{ -8.5}^{+ 14.1 }$ &$ 0.40 _{ -0.03}^{+ 0.06 }$ &$ 9.6 _{ -0.7}^{+ 2.0}$\\
& 2.3 & 1.00 & $ 29.9 _{ -6.4}^{+ 7.5 }$ &$ 74.5 _{ -3.9}^{+ 5.0 }$ &$ 0.43 _{ -0.07}^{+ 0.08 }$ &$ 10.8 _{ -0.4}^{+ 1.3}$\\
 Draco II & 1.1 & 1.00 & $ 20.9 _{ -2.1}^{+ 2.0 }$ &$ 129.2 _{ -49.3}^{+ 134.0 }$ &$ 0.72 _{ -0.10}^{+ 0.12 }$ &$ 9.3 _{ -1.7}^{+ 2.8}$\\
& 2.3 & 1.00 & $ 20.3 _{ -2.2}^{+ 1.6 }$ &$ 66.8 _{ -16.3}^{+ 27.9 }$ &$ 0.53 _{ -0.06}^{+ 0.09 }$ &$ 11.6 _{ -1.1}^{+ 0.7}$\\
 Tucana II & 1.1 & 0.80 & $ 37.7 _{ -15.9}^{+ 56.0 }$ &$ 234.2 _{ -128.3}^{+ 316.8 }$ &$ 0.72 _{ -0.08}^{+ 0.07 }$ &$ 9.4 _{ -8.8}^{+ 2.4}$\\
& 2.3 & 1.00 & $ 33.5 _{ -14.3}^{+ 16.7 }$ &$ 112.4 _{ -37.6}^{+ 120.5 }$ &$ 0.54 _{ -0.01}^{+ 0.11 }$ &$ 10.1 _{ -2.7}^{+ 2.1}$\\
Tucana III & 1.1 & 1.00 & $ 3.3 _{ -0.3}^{+ 0.6 }$ &$ 46.8 _{ -4.8}^{+ 5.6 }$ &$ 0.87 _{ -0.02}^{+ 0.01 }$ &$ 11.6 _{ -0.2}^{+ 1.4}$\\
& 2.3 & 1.00 & $ 3.1 _{ -0.3}^{+ 0.6 }$ &$ 38.8 _{ -3.5}^{+ 3.7 }$ &$ 0.85 _{ -0.02}^{+ 0.01 }$ &$ 12.0 _{ -0.1}^{+ 1.0}$\\
        \hline
	\end{tabular*}
\end{table*}

\clearpage

\onecolumn
\begin{center}
\begin{small}
\begin{longtable}{lccllll}

\caption{Full kinematic data for GCs in M1 for the upper and lower mass bounds in \citet{Watkins19}.
The mass of the bulge is small and so has been neglected.
Uncertainties are determined from the most extreme examples of the bound orbits. AM 4 is the only 
GC to be unbound for the default values in \citet{Baumgardt18}, so we have given the value as the average 
of both extremes.}
\label{full_gc} \\

\hline \multicolumn{1}{l}{GC} & \multicolumn{1}{c}{$M_{\rm h}+M_{\rm d} (\times10^{12} {\rm M_\odot})$} &
\multicolumn{1}{c}{$f_b$} & \multicolumn{1}{c}{ pericentre (kpc)} &
\multicolumn{1}{c}{apocentre (kpc)} & \multicolumn{1}{c}{eccentricity} & 
\multicolumn{1}{c}{$t_{\rm infall}$ (Gyr)} \\ \hline
\endfirsthead

\multicolumn{3}{c}%
{{\bfseries \tablename\ \thetable{} -- continued from previous page}} \\
\hline \multicolumn{1}{l}{GC} & \multicolumn{1}{c}{$M_{\rm h}+M_{\rm d} (\times10^{12} {\rm M_\odot})$} &
\multicolumn{1}{c}{$f_b$} & \multicolumn{1}{c}{ pericentre (kpc)} &
\multicolumn{1}{c}{apocentre (kpc)} & \multicolumn{1}{c}{eccentricity} & 
\multicolumn{1}{c}{$t_{\rm infall}$ (Gyr)} \\ \hline
\endhead

\hline \multicolumn{3}{|r|}{{Continued on next page}} \\ \hline
\endfoot

\hline \hline
\endlastfoot

NGC 104 & 1.1 & 1.00 & $ 6.0 _{ -0.0}^{+ 0.0 }$ &$ 7.6 _{ -0.0}^{+ 0.0 }$ &$ 0.11 _{ -0.00}^{+ 0.00 }$ &$ 12.8 _{ -0.1}^{+ 0.0}$\\
& 2.3 & 1.00 & $ 5.7 _{ -0.0}^{+ 0.0 }$ &$ 7.6 _{ -0.0}^{+ 0.0 }$ &$ 0.14 _{ -0.00}^{+ 0.00 }$ &$ 12.9 _{ -0.0}^{+ 0.0}$\\
NGC 288 & 1.1 & 1.00 & $ 3.8 _{ -0.6}^{+ 0.6 }$ &$ 13.2 _{ -0.3}^{+ 0.3 }$ &$ 0.55 _{ -0.05}^{+ 0.05 }$ &$ 12.6 _{ -0.2}^{+ 0.4}$\\
& 2.3 & 1.00 & $ 3.6 _{ -0.5}^{+ 0.5 }$ &$ 13.2 _{ -0.3}^{+ 0.3 }$ &$ 0.57 _{ -0.04}^{+ 0.04 }$ &$ 12.7 _{ -0.1}^{+ 0.3}$\\
NGC 362 & 1.1 & 1.00 & $ 1.2 _{ -0.6}^{+ 0.2 }$ &$ 11.7 _{ -8.4}^{+ 22.4 }$ &$ 0.82 _{ -0.14}^{+ 0.12 }$ &$ 10.6 _{ -6.1}^{+ 2.4}$\\
& 2.3 & 1.00 & $ 1.2 _{ -0.3}^{+ 0.1 }$ &$ 12.5 _{ -7.2}^{+ 9.0 }$ &$ 0.83 _{ -0.11}^{+ 0.07 }$ &$ 12.8 _{ -3.1}^{+ 0.2}$\\
Whiting 1 & 1.1 & 1.00 & $ 18.7 _{ -5.4}^{+ 7.0 }$ &$ 58.1 _{ -11.4}^{+ 22.8 }$ &$ 0.51 _{ -0.02}^{+ 0.05 }$ &$ 11.6 _{ -1.5}^{+ 0.8}$\\
& 2.3 & 1.00 & $ 16.4 _{ -4.6}^{+ 6.5 }$ &$ 48.0 _{ -6.7}^{+ 10.6 }$ &$ 0.49 _{ -0.06}^{+ 0.07 }$ &$ 11.8 _{ -0.6}^{+ 0.8}$\\
NGC 1261 & 1.1 & 1.00 & $ 1.5 _{ -0.2}^{+ 0.6 }$ &$ 20.6 _{ -2.2}^{+ 1.8 }$ &$ 0.87 _{ -0.03}^{+ 0.01 }$ &$ 12.6 _{ -1.3}^{+ 0.4}$\\
& 2.3 & 1.00 & $ 1.5 _{ -0.2}^{+ 0.5 }$ &$ 19.8 _{ -1.9}^{+ 1.8 }$ &$ 0.86 _{ -0.03}^{+ 0.02 }$ &$ 12.7 _{ -1.8}^{+ 0.3}$\\
Pal 1 & 1.1 & 1.00 & $ 15.2 _{ -1.2}^{+ 1.2 }$ &$ 19.5 _{ -1.7}^{+ 1.9 }$ &$ 0.12 _{ -0.02}^{+ 0.02 }$ &$ 12.2 _{ -0.2}^{+ 0.1}$\\
& 2.3 & 1.00 & $ 13.5 _{ -1.1}^{+ 1.2 }$ &$ 18.3 _{ -1.3}^{+ 1.4 }$ &$ 0.15 _{ -0.02}^{+ 0.02 }$ &$ 12.4 _{ -0.1}^{+ 0.2}$\\
AM 1 & 1.1 & 0.64 & $ 95.0 _{ -86.2}^{+ 123.5 }$ &$ 164.2 _{ -48.1}^{+ 343.2 }$ &$ 0.27 _{ -0.09}^{+ 0.62 }$ &$ 7.8 _{ -7.5}^{+ 4.4}$\\
& 2.3 & 0.97 & $ 65.7 _{ -58.4}^{+ 157.4 }$ &$ 135.1 _{ -20.2}^{+ 500.9 }$ &$ 0.35 _{ -0.22}^{+ 0.56 }$ &$ 9.0 _{ -8.4}^{+ 3.6}$\\
Eridanus & 1.1 & 1.00 & $ 31.8 _{ -20.5}^{+ 38.2 }$ &$ 157.7 _{ -29.0}^{+ 117.0 }$ &$ 0.66 _{ -0.11}^{+ 0.20 }$ &$ 8.7 _{ -3.5}^{+ 3.6}$\\
& 2.3 & 1.00 & $ 25.3 _{ -15.9}^{+ 30.0 }$ &$ 124.7 _{ -15.7}^{+ 31.5 }$ &$ 0.66 _{ -0.19}^{+ 0.21 }$ &$ 10.3 _{ -1.4}^{+ 2.4}$\\
Pal 2 & 1.1 & 1.00 & $ 2.0 _{ -1.7}^{+ 4.6 }$ &$ 40.9 _{ -19.7}^{+ 29.8 }$ &$ 0.91 _{ -0.17}^{+ 0.07 }$ &$ 12.5 _{ -7.1}^{+ 0.5}$\\
& 2.3 & 1.00 & $ 1.9 _{ -1.6}^{+ 4.0 }$ &$ 39.3 _{ -12.2}^{+ 18.9 }$ &$ 0.91 _{ -0.15}^{+ 0.08 }$ &$ 12.3 _{ -2.4}^{+ 0.7}$\\
NGC 1851 & 1.1 & 1.00 & $ 0.9 _{ -0.0}^{+ 0.3 }$ &$ 19.2 _{ -4.5}^{+ 9.8 }$ &$ 0.91 _{ -0.04}^{+ 0.02 }$ &$ 12.1 _{ -3.3}^{+ 0.9}$\\
& 2.3 & 1.00 & $ 1.1 _{ -0.2}^{+ 0.1 }$ &$ 26.8 _{ -19.5}^{+ 3.5 }$ &$ 0.92 _{ -0.13}^{+ 0.01 }$ &$ 12.5 _{ -1.7}^{+ 0.5}$\\
NGC 1904 & 1.1 & 1.00 & $ 0.7 _{ -0.5}^{+ 0.5 }$ &$ 19.7 _{ -7.2}^{+ 16.8 }$ &$ 0.93 _{ -0.05}^{+ 0.05 }$ &$ 12.5 _{ -2.4}^{+ 0.5}$\\
& 2.3 & 1.00 & $ 0.6 _{ -0.2}^{+ 0.6 }$ &$ 16.3 _{ -2.7}^{+ 14.4 }$ &$ 0.93 _{ -0.04}^{+ 0.04 }$ &$ 12.9 _{ -4.1}^{+ 0.1}$\\
NGC 2298 & 1.1 & 1.00 & $ 1.8 _{ -0.9}^{+ 1.2 }$ &$ 18.2 _{ -1.4}^{+ 1.7 }$ &$ 0.82 _{ -0.08}^{+ 0.08 }$ &$ 12.7 _{ -0.5}^{+ 0.3}$\\
& 2.3 & 1.00 & $ 1.7 _{ -1.1}^{+ 1.1 }$ &$ 17.8 _{ -1.6}^{+ 1.5 }$ &$ 0.83 _{ -0.08}^{+ 0.10 }$ &$ 12.5 _{ -0.2}^{+ 0.5}$\\
NGC 2419 & 1.1 & 1.00 & $ 16.7 _{ -3.1}^{+ 4.3 }$ &$ 92.5 _{ -8.6}^{+ 8.7 }$ &$ 0.69 _{ -0.09}^{+ 0.06 }$ &$ 10.6 _{ -0.6}^{+ 1.6}$\\
& 2.3 & 1.00 & $ 13.4 _{ -2.3}^{+ 3.1 }$ &$ 91.8 _{ -8.5}^{+ 8.5 }$ &$ 0.74 _{ -0.08}^{+ 0.05 }$ &$ 12.0 _{ -1.3}^{+ 0.4}$\\
Pyxis & 1.1 & 1.00 & $ 24.2 _{ -5.9}^{+ 7.2 }$ &$ 181.5 _{ -61.2}^{+ 141.5 }$ &$ 0.76 _{ -0.03}^{+ 0.06 }$ &$ 10.3 _{ -4.6}^{+ 1.8}$\\
& 2.3 & 1.00 & $ 22.1 _{ -5.3}^{+ 5.9 }$ &$ 94.6 _{ -19.1}^{+ 28.5 }$ &$ 0.62 _{ -0.01}^{+ 0.02 }$ &$ 10.6 _{ -0.7}^{+ 1.7}$\\
NGC 2808 & 1.1 & 1.00 & $ 0.9 _{ -0.1}^{+ 0.0 }$ &$ 15.3 _{ -0.2}^{+ 0.2 }$ &$ 0.89 _{ -0.00}^{+ 0.01 }$ &$ 12.9 _{ -0.5}^{+ 0.1}$\\
& 2.3 & 1.00 & $ 0.9 _{ -0.0}^{+ 0.0 }$ &$ 14.8 _{ -0.2}^{+ 0.2 }$ &$ 0.89 _{ -0.01}^{+ 0.01 }$ &$ 12.9 _{ -0.4}^{+ 0.1}$\\
E 3 & 1.1 & 1.00 & $ 9.3 _{ -0.5}^{+ 0.5 }$ &$ 14.7 _{ -3.0}^{+ 4.3 }$ &$ 0.23 _{ -0.09}^{+ 0.10 }$ &$ 12.3 _{ -0.2}^{+ 0.3}$\\
& 2.3 & 1.00 & $ 9.2 _{ -0.5}^{+ 0.5 }$ &$ 12.7 _{ -2.3}^{+ 3.1 }$ &$ 0.16 _{ -0.07}^{+ 0.08 }$ &$ 12.7 _{ -0.2}^{+ 0.0}$\\
Pal 3 & 1.1 & 1.00 & $ 65.8 _{ -46.2}^{+ 63.0 }$ &$ 122.4 _{ -27.0}^{+ 316.9 }$ &$ 0.30 _{ -0.04}^{+ 0.41 }$ &$ 8.4 _{ -7.6}^{+ 3.3}$\\
& 2.3 & 1.00 & $ 47.7 _{ -32.1}^{+ 47.8 }$ &$ 105.3 _{ -13.8}^{+ 58.1 }$ &$ 0.38 _{ -0.17}^{+ 0.38 }$ &$ 10.1 _{ -2.0}^{+ 1.8}$\\
NGC 3201 & 1.1 & 1.00 & $ 8.3 _{ -0.1}^{+ 0.1 }$ &$ 27.2 _{ -1.5}^{+ 1.7 }$ &$ 0.53 _{ -0.02}^{+ 0.02 }$ &$ 12.5 _{ -0.6}^{+ 0.1}$\\
& 2.3 & 1.00 & $ 8.3 _{ -0.1}^{+ 0.1 }$ &$ 21.4 _{ -0.9}^{+ 1.0 }$ &$ 0.44 _{ -0.01}^{+ 0.01 }$ &$ 12.6 _{ -0.3}^{+ 0.2}$\\
Pal 4 & 1.1 & 1.00 & $ 8.9 _{ -7.1}^{+ 37.3 }$ &$ 111.0 _{ -13.1}^{+ 15.5 }$ &$ 0.85 _{ -0.47}^{+ 0.12 }$ &$ 12.5 _{ -3.8}^{+ 0.3}$\\
& 2.3 & 1.00 & $ 7.6 _{ -6.0}^{+ 26.0 }$ &$ 108.9 _{ -12.7}^{+ 12.6 }$ &$ 0.87 _{ -0.38}^{+ 0.10 }$ &$ 12.4 _{ -2.2}^{+ 0.5}$\\
Crater & 1.1 & 0.69 & $ 87.5 _{ -85.2}^{+ 184.8 }$ &$ 147.1 _{ -16.6}^{+ 326.0 }$ &$ 0.25 _{ -0.24}^{+ 0.72 }$ &$ 6.6 _{ -6.2}^{+ 6.3}$\\
& 2.3 & 0.90 & $ 56.3 _{ -54.3}^{+ 169.2 }$ &$ 145.6 _{ -15.2}^{+ 435.2 }$ &$ 0.44 _{ -0.43}^{+ 0.53 }$ &$ 8.8 _{ -8.2}^{+ 4.0}$\\
NGC 4147 & 1.1 & 1.00 & $ 1.9 _{ -0.6}^{+ 0.9 }$ &$ 26.0 _{ -4.2}^{+ 7.5 }$ &$ 0.86 _{ -0.08}^{+ 0.06 }$ &$ 12.2 _{ -0.6}^{+ 0.8}$\\
& 2.3 & 1.00 & $ 1.8 _{ -0.3}^{+ 1.0 }$ &$ 24.3 _{ -2.2}^{+ 2.9 }$ &$ 0.87 _{ -0.08}^{+ 0.03 }$ &$ 12.9 _{ -1.0}^{+ 0.1}$\\
NGC 4372 & 1.1 & 1.00 & $ 3.1 _{ -0.1}^{+ 0.1 }$ &$ 7.4 _{ -0.1}^{+ 0.1 }$ &$ 0.41 _{ -0.01}^{+ 0.01 }$ &$ 13.0 _{ -0.2}^{+ 0.0}$\\
& 2.3 & 1.00 & $ 3.0 _{ -0.1}^{+ 0.1 }$ &$ 7.4 _{ -0.1}^{+ 0.1 }$ &$ 0.42 _{ -0.01}^{+ 0.01 }$ &$ 12.8 _{ -0.0}^{+ 0.2}$\\
Rup 106 & 1.1 & 1.00 & $ 4.9 _{ -0.7}^{+ 0.9 }$ &$ 38.9 _{ -4.8}^{+ 5.3 }$ &$ 0.77 _{ -0.01}^{+ 0.01 }$ &$ 12.8 _{ -1.1}^{+ 0.1}$\\
& 2.3 & 1.00 & $ 4.7 _{ -0.7}^{+ 0.9 }$ &$ 32.4 _{ -3.5}^{+ 3.7 }$ &$ 0.75 _{ -0.01}^{+ 0.01 }$ &$ 12.5 _{ -0.3}^{+ 0.4}$\\
NGC 4590 & 1.1 & 1.00 & $ 9.0 _{ -0.4}^{+ 0.4 }$ &$ 34.1 _{ -3.5}^{+ 4.2 }$ &$ 0.58 _{ -0.02}^{+ 0.02 }$ &$ 11.8 _{ -0.1}^{+ 0.8}$\\
& 2.3 & 1.00 & $ 8.9 _{ -0.4}^{+ 0.5 }$ &$ 25.7 _{ -2.1}^{+ 2.4 }$ &$ 0.49 _{ -0.02}^{+ 0.02 }$ &$ 12.5 _{ -0.4}^{+ 0.3}$\\
NGC 4833 & 1.1 & 1.00 & $ 0.6 _{ -0.2}^{+ 0.2 }$ &$ 7.8 _{ -0.6}^{+ 0.3 }$ &$ 0.86 _{ -0.03}^{+ 0.05 }$ &$ 12.8 _{ -0.5}^{+ 0.2}$\\
& 2.3 & 1.00 & $ 0.6 _{ -0.1}^{+ 0.2 }$ &$ 7.7 _{ -0.7}^{+ 0.3 }$ &$ 0.86 _{ -0.03}^{+ 0.02 }$ &$ 13.0 _{ -0.5}^{+ 0.0}$\\
NGC 5024 & 1.1 & 1.00 & $ 10.0 _{ -0.3}^{+ 0.3 }$ &$ 23.2 _{ -1.8}^{+ 1.9 }$ &$ 0.40 _{ -0.03}^{+ 0.03 }$ &$ 12.1 _{ -0.2}^{+ 0.5}$\\
& 2.3 & 1.00 & $ 9.1 _{ -0.2}^{+ 0.2 }$ &$ 21.6 _{ -1.7}^{+ 1.8 }$ &$ 0.41 _{ -0.03}^{+ 0.03 }$ &$ 12.7 _{ -0.5}^{+ 0.1}$\\
NGC 5053 & 1.1 & 1.00 & $ 11.7 _{ -0.2}^{+ 0.1 }$ &$ 17.7 _{ -1.5}^{+ 1.6 }$ &$ 0.20 _{ -0.05}^{+ 0.05 }$ &$ 12.3 _{ -0.2}^{+ 0.1}$\\
& 2.3 & 1.00 & $ 10.2 _{ -0.2}^{+ 0.1 }$ &$ 17.7 _{ -1.5}^{+ 1.6 }$ &$ 0.27 _{ -0.05}^{+ 0.05 }$ &$ 12.5 _{ -0.0}^{+ 0.2}$\\
NGC 5139 & 1.1 & 1.00 & $ 1.8 _{ -0.0}^{+ 0.0 }$ &$ 7.0 _{ -0.0}^{+ 0.0 }$ &$ 0.59 _{ -0.01}^{+ 0.01 }$ &$ 12.9 _{ -0.2}^{+ 0.1}$\\
& 2.3 & 1.00 & $ 1.7 _{ -0.0}^{+ 0.0 }$ &$ 7.0 _{ -0.0}^{+ 0.0 }$ &$ 0.60 _{ -0.01}^{+ 0.01 }$ &$ 12.9 _{ -0.0}^{+ 0.1}$\\
NGC 5272 & 1.1 & 1.00 & $ 6.0 _{ -0.1}^{+ 0.1 }$ &$ 16.1 _{ -0.5}^{+ 0.5 }$ &$ 0.46 _{ -0.02}^{+ 0.02 }$ &$ 12.6 _{ -0.3}^{+ 0.2}$\\
& 2.3 & 1.00 & $ 5.7 _{ -0.1}^{+ 0.1 }$ &$ 15.1 _{ -0.4}^{+ 0.4 }$ &$ 0.45 _{ -0.02}^{+ 0.02 }$ &$ 12.7 _{ -0.2}^{+ 0.2}$\\
NGC 5286 & 1.1 & 1.00 & $ 1.0 _{ -0.3}^{+ 0.4 }$ &$ 13.9 _{ -1.1}^{+ 0.9 }$ &$ 0.87 _{ -0.04}^{+ 0.03 }$ &$ 13.0 _{ -0.7}^{+ 0.0}$\\
& 2.3 & 1.00 & $ 1.1 _{ -0.2}^{+ 0.3 }$ &$ 13.0 _{ -1.1}^{+ 0.9 }$ &$ 0.85 _{ -0.03}^{+ 0.03 }$ &$ 12.7 _{ -0.3}^{+ 0.3}$\\
AM 4 & 1.1 & 0.37 & $ 36.9\pm 12.7$ & $ 421.2\pm 372.4$ & $ 0.63\pm 0.30$ & $ 6.1 _{ -5.7}^{+ 5.7}$\\
& 2.3 & 0.72 & $ 28.2 _{ -4.1}^{+ 22.3 }$ &$ 207.7 _{ -174.0}^{+ 890.9 }$ &$ 0.76 _{ -0.61}^{+ 0.19 }$ &$ 9.1 _{ -8.7}^{+ 2.9}$\\
NGC 5466 & 1.1 & 1.00 & $ 7.9 _{ -2.5}^{+ 2.7 }$ &$ 72.3 _{ -23.7}^{+ 48.5 }$ &$ 0.80 _{ -0.01}^{+ 0.04 }$ &$ 11.8 _{ -1.9}^{+ 0.9}$\\
& 2.3 & 1.00 & $ 7.7 _{ -2.5}^{+ 2.6 }$ &$ 47.7 _{ -11.1}^{+ 17.4 }$ &$ 0.72 _{ -0.00}^{+ 0.03 }$ &$ 11.8 _{ -0.3}^{+ 1.1}$\\
NGC 5634 & 1.1 & 1.00 & $ 3.9 _{ -1.5}^{+ 2.7 }$ &$ 23.6 _{ -2.9}^{+ 2.8 }$ &$ 0.71 _{ -0.12}^{+ 0.08 }$ &$ 12.3 _{ -0.3}^{+ 0.7}$\\
& 2.3 & 1.00 & $ 3.7 _{ -1.3}^{+ 2.3 }$ &$ 23.4 _{ -2.8}^{+ 2.8 }$ &$ 0.73 _{ -0.10}^{+ 0.07 }$ &$ 12.6 _{ -0.4}^{+ 0.3}$\\
NGC 5694 & 1.1 & 1.00 & $ 3.8 _{ -0.6}^{+ 1.4 }$ &$ 74.5 _{ -10.8}^{+ 13.1 }$ &$ 0.90 _{ -0.02}^{+ 0.02 }$ &$ 11.5 _{ -0.9}^{+ 1.5}$\\
& 2.3 & 1.00 & $ 3.6 _{ -0.6}^{+ 1.3 }$ &$ 56.5 _{ -6.8}^{+ 7.7 }$ &$ 0.88 _{ -0.02}^{+ 0.01 }$ &$ 11.7 _{ -0.2}^{+ 1.3}$\\
IC 4499 & 1.1 & 1.00 & $ 6.4 _{ -1.4}^{+ 1.4 }$ &$ 29.9 _{ -4.1}^{+ 4.5 }$ &$ 0.65 _{ -0.02}^{+ 0.03 }$ &$ 12.4 _{ -0.4}^{+ 0.5}$\\
& 2.3 & 1.00 & $ 6.1 _{ -1.3}^{+ 1.4 }$ &$ 25.5 _{ -2.9}^{+ 3.1 }$ &$ 0.62 _{ -0.03}^{+ 0.04 }$ &$ 12.6 _{ -0.4}^{+ 0.4}$\\
NGC 5824 & 1.1 & 1.00 & $ 14.9 _{ -5.8}^{+ 6.5 }$ &$ 37.4 _{ -8.0}^{+ 14.7 }$ &$ 0.43 _{ -0.02}^{+ 0.10 }$ &$ 11.6 _{ -0.7}^{+ 0.9}$\\
& 2.3 & 1.00 & $ 13.3 _{ -5.1}^{+ 6.2 }$ &$ 32.5 _{ -5.4}^{+ 7.8 }$ &$ 0.42 _{ -0.07}^{+ 0.12 }$ &$ 12.3 _{ -0.6}^{+ 0.4}$\\
Pal 5 & 1.1 & 1.00 & $ 13.5 _{ -5.8}^{+ 6.2 }$ &$ 19.2 _{ -2.8}^{+ 7.5 }$ &$ 0.18 _{ -0.07}^{+ 0.19 }$ &$ 12.2 _{ -0.4}^{+ 0.5}$\\
& 2.3 & 1.00 & $ 11.7 _{ -4.8}^{+ 6.6 }$ &$ 18.9 _{ -2.5}^{+ 3.6 }$ &$ 0.23 _{ -0.13}^{+ 0.17 }$ &$ 12.4 _{ -0.1}^{+ 0.5}$\\
NGC 5897 & 1.1 & 1.00 & $ 2.9 _{ -1.0}^{+ 1.2 }$ &$ 9.2 _{ -1.5}^{+ 1.9 }$ &$ 0.51 _{ -0.06}^{+ 0.08 }$ &$ 12.7 _{ -0.2}^{+ 0.3}$\\
& 2.3 & 1.00 & $ 2.9 _{ -1.0}^{+ 1.2 }$ &$ 8.9 _{ -1.4}^{+ 1.7 }$ &$ 0.51 _{ -0.07}^{+ 0.09 }$ &$ 13.0 _{ -0.2}^{+ 0.0}$\\
NGC 5904 & 1.1 & 1.00 & $ 2.8 _{ -0.1}^{+ 0.2 }$ &$ 30.0 _{ -1.7}^{+ 2.3 }$ &$ 0.83 _{ -0.01}^{+ 0.00 }$ &$ 12.2 _{ -0.3}^{+ 0.8}$\\
& 2.3 & 1.00 & $ 2.8 _{ -0.1}^{+ 0.1 }$ &$ 23.6 _{ -1.1}^{+ 1.3 }$ &$ 0.78 _{ -0.01}^{+ 0.01 }$ &$ 12.8 _{ -0.5}^{+ 0.2}$\\
NGC 5927 & 1.1 & 1.00 & $ 4.3 _{ -0.0}^{+ 0.0 }$ &$ 5.5 _{ -0.2}^{+ 0.2 }$ &$ 0.13 _{ -0.02}^{+ 0.02 }$ &$ 12.9 _{ -0.0}^{+ 0.1}$\\
& 2.3 & 1.00 & $ 4.2 _{ -0.0}^{+ 0.0 }$ &$ 5.4 _{ -0.2}^{+ 0.2 }$ &$ 0.12 _{ -0.02}^{+ 0.02 }$ &$ 12.9 _{ -0.0}^{+ 0.1}$\\
NGC 5946 & 1.1 & 1.00 & $ 0.8 _{ -0.2}^{+ 0.5 }$ &$ 5.9 _{ -1.0}^{+ 1.1 }$ &$ 0.75 _{ -0.08}^{+ 0.01 }$ &$ 13.0 _{ -2.3}^{+ 0.0}$\\
& 2.3 & 1.00 & $ 0.8 _{ -0.1}^{+ 0.5 }$ &$ 5.9 _{ -0.9}^{+ 0.9 }$ &$ 0.77 _{ -0.09}^{+ 0.02 }$ &$ 13.0 _{ -1.8}^{+ 0.0}$\\
NGC 5986 & 1.1 & 1.00 & $ 0.5 _{ -0.1}^{+ 0.7 }$ &$ 5.5 _{ -0.9}^{+ 0.4 }$ &$ 0.84 _{ -0.20}^{+ 0.04 }$ &$ 13.0 _{ -1.8}^{+ 0.0}$\\
& 2.3 & 1.00 & $ 0.8 _{ -0.4}^{+ 0.4 }$ &$ 5.0 _{ -0.4}^{+ 0.9 }$ &$ 0.73 _{ -0.08}^{+ 0.13 }$ &$ 13.0 _{ -2.6}^{+ 0.0}$\\
FSR 1716 & 1.1 & 1.00 & $ 2.5 _{ -0.3}^{+ 0.5 }$ &$ 5.0 _{ -0.2}^{+ 0.5 }$ &$ 0.34 _{ -0.05}^{+ 0.04 }$ &$ 12.9 _{ -0.1}^{+ 0.1}$\\
& 2.3 & 1.00 & $ 2.4 _{ -0.3}^{+ 0.5 }$ &$ 4.9 _{ -0.2}^{+ 0.5 }$ &$ 0.34 _{ -0.05}^{+ 0.05 }$ &$ 13.0 _{ -0.0}^{+ 0.0}$\\
Pal 14 & 1.1 & 1.00 & $ 1.1 _{ -0.6}^{+ 7.2 }$ &$ 82.3 _{ -31.1}^{+ 41.2 }$ &$ 0.97 _{ -0.10}^{+ 0.02 }$ &$ 11.3 _{ -3.3}^{+ 1.7}$\\
& 2.3 & 1.00 & $ 1.3 _{ -0.9}^{+ 6.0 }$ &$ 79.3 _{ -22.1}^{+ 39.3 }$ &$ 0.97 _{ -0.10}^{+ 0.03 }$ &$ 12.0 _{ -1.8}^{+ 1.0}$\\
Lynga 7 & 1.1 & 1.00 & $ 1.9 _{ -0.3}^{+ 0.5 }$ &$ 4.6 _{ -0.2}^{+ 0.3 }$ &$ 0.42 _{ -0.06}^{+ 0.05 }$ &$ 12.9 _{ -0.0}^{+ 0.1}$\\
& 2.3 & 1.00 & $ 1.8 _{ -0.3}^{+ 0.4 }$ &$ 4.6 _{ -0.2}^{+ 0.3 }$ &$ 0.43 _{ -0.06}^{+ 0.05 }$ &$ 13.0 _{ -0.0}^{+ 0.0}$\\
NGC 6093 & 1.1 & 1.00 & $ 0.6 _{ -0.3}^{+ 0.5 }$ &$ 2.5 _{ -1.3}^{+ 8.1 }$ &$ 0.60 _{ -0.07}^{+ 0.26 }$ &$ 12.9 _{ -4.0}^{+ 0.1}$\\
& 2.3 & 1.00 & $ 0.9 _{ -0.6}^{+ 0.2 }$ &$ 4.7 _{ -3.6}^{+ 6.1 }$ &$ 0.68 _{ -0.14}^{+ 0.17 }$ &$ 12.9 _{ -3.0}^{+ 0.1}$\\
NGC 6121 & 1.1 & 1.00 & $ 0.5 _{ -0.1}^{+ 0.1 }$ &$ 6.5 _{ -0.1}^{+ 0.1 }$ &$ 0.85 _{ -0.02}^{+ 0.02 }$ &$ 13.0 _{ -0.4}^{+ 0.0}$\\
& 2.3 & 1.00 & $ 0.5 _{ -0.1}^{+ 0.1 }$ &$ 6.6 _{ -0.1}^{+ 0.0 }$ &$ 0.85 _{ -0.02}^{+ 0.02 }$ &$ 12.9 _{ -0.3}^{+ 0.1}$\\
NGC 6101 & 1.1 & 1.00 & $ 11.5 _{ -1.2}^{+ 1.2 }$ &$ 57.5 _{ -14.5}^{+ 20.3 }$ &$ 0.67 _{ -0.05}^{+ 0.05 }$ &$ 11.5 _{ -0.7}^{+ 1.0}$\\
& 2.3 & 1.00 & $ 11.4 _{ -1.2}^{+ 1.2 }$ &$ 37.4 _{ -7.1}^{+ 8.6 }$ &$ 0.53 _{ -0.04}^{+ 0.04 }$ &$ 12.4 _{ -0.4}^{+ 0.3}$\\
NGC 6144 & 1.1 & 1.00 & $ 2.1 _{ -0.7}^{+ 0.3 }$ &$ 3.9 _{ -0.1}^{+ 0.3 }$ &$ 0.30 _{ -0.05}^{+ 0.18 }$ &$ 13.0 _{ -0.1}^{+ 0.0}$\\
& 2.3 & 1.00 & $ 2.2 _{ -0.8}^{+ 0.3 }$ &$ 3.7 _{ -0.0}^{+ 0.4 }$ &$ 0.26 _{ -0.04}^{+ 0.21 }$ &$ 13.0 _{ -0.0}^{+ 0.0}$\\
NGC 6139 & 1.1 & 1.00 & $ 1.3 _{ -0.3}^{+ 0.6 }$ &$ 3.4 _{ -0.5}^{+ 0.6 }$ &$ 0.45 _{ -0.08}^{+ 0.07 }$ &$ 13.0 _{ -0.1}^{+ 0.0}$\\
& 2.3 & 1.00 & $ 1.3 _{ -0.3}^{+ 0.5 }$ &$ 3.4 _{ -0.5}^{+ 0.6 }$ &$ 0.46 _{ -0.08}^{+ 0.06 }$ &$ 13.0 _{ -0.0}^{+ 0.0}$\\
Ter 3 & 1.1 & 1.00 & $ 2.3 _{ -0.1}^{+ 0.1 }$ &$ 3.6 _{ -0.1}^{+ 0.4 }$ &$ 0.22 _{ -0.03}^{+ 0.07 }$ &$ 13.0 _{ -0.0}^{+ 0.0}$\\
& 2.3 & 1.00 & $ 2.3 _{ -0.1}^{+ 0.2 }$ &$ 3.5 _{ -0.1}^{+ 0.4 }$ &$ 0.20 _{ -0.04}^{+ 0.07 }$ &$ 13.0 _{ 0.0}^{+ 0.0}$\\
NGC 6171 & 1.1 & 1.00 & $ 1.5 _{ -0.1}^{+ 0.1 }$ &$ 3.7 _{ -0.1}^{+ 0.1 }$ &$ 0.42 _{ -0.02}^{+ 0.01 }$ &$ 13.0 _{ -0.1}^{+ 0.0}$\\
& 2.3 & 1.00 & $ 1.5 _{ -0.1}^{+ 0.1 }$ &$ 3.7 _{ -0.2}^{+ 0.1 }$ &$ 0.42 _{ -0.04}^{+ 0.01 }$ &$ 13.0 _{ -0.0}^{+ 0.0}$\\
ESO 452-SC11 & 1.1 & 1.00 & $ 0.8 _{ -0.5}^{+ 0.4 }$ &$ 3.0 _{ -0.7}^{+ 3.5 }$ &$ 0.60 _{ -0.13}^{+ 0.33 }$ &$ 13.0 _{ -5.8}^{+ 0.0}$\\
& 2.3 & 1.00 & $ 0.7 _{ -0.5}^{+ 0.4 }$ &$ 3.0 _{ -0.9}^{+ 2.1 }$ &$ 0.62 _{ -0.14}^{+ 0.27 }$ &$ 13.0 _{ -2.2}^{+ 0.0}$\\
NGC 6205 & 1.1 & 1.00 & $ 1.7 _{ -0.1}^{+ 0.0 }$ &$ 8.5 _{ -0.2}^{+ 0.2 }$ &$ 0.67 _{ -0.01}^{+ 0.01 }$ &$ 12.7 _{ -0.1}^{+ 0.3}$\\
& 2.3 & 1.00 & $ 1.6 _{ -0.1}^{+ 0.0 }$ &$ 8.5 _{ -0.2}^{+ 0.2 }$ &$ 0.68 _{ -0.01}^{+ 0.01 }$ &$ 12.9 _{ -0.2}^{+ 0.1}$\\
NGC 6229 & 1.1 & 1.00 & $ 1.8 _{ -0.7}^{+ 2.3 }$ &$ 31.3 _{ -12.3}^{+ 21.6 }$ &$ 0.89 _{ -0.11}^{+ 0.06 }$ &$ 12.5 _{ -4.7}^{+ 0.5}$\\
& 2.3 & 1.00 & $ 1.6 _{ -0.7}^{+ 2.2 }$ &$ 31.1 _{ -12.4}^{+ 12.8 }$ &$ 0.90 _{ -0.10}^{+ 0.04 }$ &$ 12.7 _{ -2.4}^{+ 0.3}$\\
NGC 6218 & 1.1 & 1.00 & $ 2.7 _{ -0.1}^{+ 0.1 }$ &$ 5.0 _{ -0.1}^{+ 0.1 }$ &$ 0.29 _{ -0.01}^{+ 0.01 }$ &$ 12.9 _{ -0.0}^{+ 0.1}$\\
& 2.3 & 1.00 & $ 2.7 _{ -0.1}^{+ 0.1 }$ &$ 5.0 _{ -0.1}^{+ 0.1 }$ &$ 0.30 _{ -0.01}^{+ 0.01 }$ &$ 13.0 _{ -0.0}^{+ 0.0}$\\
FSR 1735 & 1.1 & 1.00 & $ 1.3 _{ -0.4}^{+ 0.2 }$ &$ 5.1 _{ -1.1}^{+ 1.4 }$ &$ 0.60 _{ -0.12}^{+ 0.16 }$ &$ 13.0 _{ -0.5}^{+ 0.0}$\\
& 2.3 & 1.00 & $ 1.2 _{ -0.5}^{+ 0.3 }$ &$ 5.1 _{ -1.0}^{+ 1.4 }$ &$ 0.61 _{ -0.13}^{+ 0.18 }$ &$ 13.0 _{ -0.4}^{+ 0.0}$\\
NGC 6235 & 1.1 & 1.00 & $ 5.1 _{ -1.5}^{+ 1.5 }$ &$ 16.6 _{ -7.7}^{+ 18.4 }$ &$ 0.53 _{ -0.11}^{+ 0.15 }$ &$ 12.7 _{ -1.0}^{+ 0.2}$\\
& 2.3 & 1.00 & $ 5.1 _{ -1.4}^{+ 1.5 }$ &$ 14.3 _{ -6.0}^{+ 11.6 }$ &$ 0.48 _{ -0.09}^{+ 0.12 }$ &$ 12.6 _{ -0.4}^{+ 0.4}$\\
NGC 6254 & 1.1 & 1.00 & $ 2.2 _{ -0.0}^{+ 0.1 }$ &$ 4.9 _{ -0.1}^{+ 0.1 }$ &$ 0.37 _{ -0.01}^{+ 0.00 }$ &$ 13.0 _{ -0.1}^{+ 0.0}$\\
& 2.3 & 1.00 & $ 2.2 _{ -0.1}^{+ 0.1 }$ &$ 4.8 _{ -0.1}^{+ 0.1 }$ &$ 0.38 _{ -0.00}^{+ 0.00 }$ &$ 13.0 _{ -0.0}^{+ 0.0}$\\
NGC 6256 & 1.1 & 1.00 & $ 2.5 _{ -0.5}^{+ 0.4 }$ &$ 2.8 _{ -0.3}^{+ 0.6 }$ &$ 0.06 _{ -0.01}^{+ 0.08 }$ &$ 13.0 _{ 0.0}^{+ 0.0}$\\
& 2.3 & 1.00 & $ 2.5 _{ -0.5}^{+ 0.3 }$ &$ 2.7 _{ -0.3}^{+ 0.6 }$ &$ 0.04 _{ -0.01}^{+ 0.09 }$ &$ 13.0 _{ 0.0}^{+ 0.0}$\\
Pal 15 & 1.1 & 1.00 & $ 0.9 _{ -0.2}^{+ 2.3 }$ &$ 49.5 _{ -7.0}^{+ 21.7 }$ &$ 0.96 _{ -0.09}^{+ 0.01 }$ &$ 11.8 _{ -2.2}^{+ 1.2}$\\
& 2.3 & 1.00 & $ 1.0 _{ -0.3}^{+ 2.0 }$ &$ 52.1 _{ -14.9}^{+ 10.2 }$ &$ 0.96 _{ -0.09}^{+ 0.01 }$ &$ 12.9 _{ -2.4}^{+ 0.1}$\\
NGC 6266 & 1.1 & 1.00 & $ 1.5 _{ -0.1}^{+ 0.2 }$ &$ 2.3 _{ -0.2}^{+ 0.2 }$ &$ 0.21 _{ -0.09}^{+ 0.06 }$ &$ 13.0 _{ 0.0}^{+ 0.0}$\\
& 2.3 & 1.00 & $ 1.4 _{ -0.1}^{+ 0.1 }$ &$ 2.4 _{ -0.0}^{+ 0.0 }$ &$ 0.25 _{ -0.03}^{+ 0.02 }$ &$ 13.0 _{ 0.0}^{+ 0.0}$\\
NGC 6273 & 1.1 & 1.00 & $ 0.8 _{ -0.0}^{+ 0.3 }$ &$ 4.6 _{ -0.1}^{+ 0.3 }$ &$ 0.70 _{ -0.07}^{+ 0.00 }$ &$ 13.0 _{ -1.8}^{+ 0.0}$\\
& 2.3 & 1.00 & $ 0.8 _{ -0.1}^{+ 0.2 }$ &$ 4.6 _{ -0.5}^{+ 0.2 }$ &$ 0.69 _{ -0.06}^{+ 0.01 }$ &$ 13.0 _{ -1.8}^{+ 0.0}$\\
NGC 6284 & 1.1 & 1.00 & $ 1.1 _{ -0.7}^{+ 0.6 }$ &$ 9.0 _{ -5.6}^{+ 6.0 }$ &$ 0.78 _{ -0.10}^{+ 0.08 }$ &$ 11.8 _{ -3.2}^{+ 1.2}$\\
& 2.3 & 1.00 & $ 1.1 _{ -0.7}^{+ 0.5 }$ &$ 11.0 _{ -9.2}^{+ 3.6 }$ &$ 0.81 _{ -0.17}^{+ 0.10 }$ &$ 13.0 _{ -4.2}^{+ 0.0}$\\
NGC 6287 & 1.1 & 1.00 & $ 0.9 _{ -0.6}^{+ 0.4 }$ &$ 5.5 _{ -3.9}^{+ 15.4 }$ &$ 0.73 _{ -0.11}^{+ 0.20 }$ &$ 11.4 _{ -7.2}^{+ 1.6}$\\
& 2.3 & 1.00 & $ 1.1 _{ -0.8}^{+ 0.2 }$ &$ 6.6 _{ -5.1}^{+ 7.8 }$ &$ 0.72 _{ -0.07}^{+ 0.18 }$ &$ 12.9 _{ -3.7}^{+ 0.1}$\\
NGC 6293 & 1.1 & 1.00 & $ 0.8 _{ -0.4}^{+ 0.2 }$ &$ 3.0 _{ -1.4}^{+ 3.6 }$ &$ 0.58 _{ -0.04}^{+ 0.20 }$ &$ 13.0 _{ -3.1}^{+ 0.0}$\\
& 2.3 & 1.00 & $ 0.8 _{ -0.4}^{+ 0.1 }$ &$ 2.8 _{ -1.2}^{+ 2.1 }$ &$ 0.55 _{ -0.01}^{+ 0.15 }$ &$ 13.0 _{ -1.7}^{+ 0.0}$\\
NGC 6304 & 1.1 & 1.00 & $ 2.0 _{ -0.2}^{+ 0.2 }$ &$ 3.5 _{ -0.4}^{+ 0.4 }$ &$ 0.26 _{ -0.01}^{+ 0.01 }$ &$ 13.0 _{ -0.0}^{+ 0.0}$\\
& 2.3 & 1.00 & $ 2.0 _{ -0.2}^{+ 0.2 }$ &$ 3.4 _{ -0.4}^{+ 0.4 }$ &$ 0.26 _{ -0.01}^{+ 0.01 }$ &$ 13.0 _{ 0.0}^{+ 0.0}$\\
NGC 6316 & 1.1 & 1.00 & $ 1.5 _{ -0.7}^{+ 1.0 }$ &$ 4.3 _{ -1.3}^{+ 1.6 }$ &$ 0.47 _{ -0.07}^{+ 0.10 }$ &$ 13.0 _{ -0.3}^{+ 0.0}$\\
& 2.3 & 1.00 & $ 1.5 _{ -0.7}^{+ 0.9 }$ &$ 4.3 _{ -1.3}^{+ 1.6 }$ &$ 0.48 _{ -0.07}^{+ 0.10 }$ &$ 13.0 _{ -0.3}^{+ 0.0}$\\
NGC 6341 & 1.1 & 1.00 & $ 1.3 _{ -0.1}^{+ 0.1 }$ &$ 10.5 _{ -3.7}^{+ 4.9 }$ &$ 0.79 _{ -0.08}^{+ 0.06 }$ &$ 12.8 _{ -3.2}^{+ 0.2}$\\
& 2.3 & 1.00 & $ 1.3 _{ -0.1}^{+ 0.1 }$ &$ 11.4 _{ -3.4}^{+ 4.0 }$ &$ 0.80 _{ -0.05}^{+ 0.05 }$ &$ 12.8 _{ -0.5}^{+ 0.2}$\\
NGC 6325 & 1.1 & 1.00 & $ 0.8 _{ -0.6}^{+ 0.7 }$ &$ 2.3 _{ -0.5}^{+ 4.8 }$ &$ 0.48 _{ -0.38}^{+ 0.44 }$ &$ 13.0 _{ -5.6}^{+ 0.0}$\\
& 2.3 & 1.00 & $ 0.8 _{ -0.5}^{+ 0.7 }$ &$ 2.3 _{ -0.5}^{+ 3.0 }$ &$ 0.46 _{ -0.36}^{+ 0.40 }$ &$ 13.0 _{ -2.9}^{+ 0.0}$\\
NGC 6333 & 1.1 & 1.00 & $ 1.2 _{ -0.2}^{+ 0.1 }$ &$ 8.3 _{ -0.1}^{+ 0.9 }$ &$ 0.74 _{ -0.01}^{+ 0.04 }$ &$ 12.6 _{ -0.1}^{+ 0.4}$\\
& 2.3 & 1.00 & $ 1.3 _{ -0.1}^{+ 0.0 }$ &$ 7.7 _{ -0.1}^{+ 0.8 }$ &$ 0.72 _{ -0.01}^{+ 0.04 }$ &$ 12.9 _{ -0.2}^{+ 0.1}$\\
NGC 6342 & 1.1 & 1.00 & $ 1.3 _{ -0.4}^{+ 0.3 }$ &$ 1.9 _{ -0.0}^{+ 0.5 }$ &$ 0.21 _{ -0.02}^{+ 0.24 }$ &$ 13.0 _{ -0.1}^{+ 0.0}$\\
& 2.3 & 1.00 & $ 1.2 _{ -0.4}^{+ 0.4 }$ &$ 1.9 _{ -0.0}^{+ 0.5 }$ &$ 0.20 _{ -0.03}^{+ 0.24 }$ &$ 13.0 _{ -0.1}^{+ 0.0}$\\
NGC 6356 & 1.1 & 1.00 & $ 2.9 _{ -1.3}^{+ 2.0 }$ &$ 7.9 _{ -1.7}^{+ 2.0 }$ &$ 0.46 _{ -0.13}^{+ 0.12 }$ &$ 12.7 _{ -0.1}^{+ 0.3}$\\
& 2.3 & 1.00 & $ 2.8 _{ -1.2}^{+ 1.9 }$ &$ 7.9 _{ -1.6}^{+ 1.9 }$ &$ 0.47 _{ -0.13}^{+ 0.12 }$ &$ 12.9 _{ -0.2}^{+ 0.1}$\\
NGC 6355 & 1.1 & 1.00 & $ 0.5 _{ -0.3}^{+ 0.5 }$ &$ 2.7 _{ -1.8}^{+ 6.9 }$ &$ 0.71 _{ -0.10}^{+ 0.24 }$ &$ 12.5 _{ -6.6}^{+ 0.5}$\\
& 2.3 & 1.00 & $ 0.5 _{ -0.4}^{+ 0.3 }$ &$ 3.8 _{ -2.7}^{+ 5.0 }$ &$ 0.75 _{ -0.17}^{+ 0.19 }$ &$ 12.5 _{ -3.0}^{+ 0.5}$\\
NGC 6352 & 1.1 & 1.00 & $ 3.2 _{ -0.1}^{+ 0.2 }$ &$ 4.0 _{ -0.5}^{+ 0.6 }$ &$ 0.11 _{ -0.04}^{+ 0.04 }$ &$ 13.0 _{ -0.1}^{+ 0.0}$\\
& 2.3 & 1.00 & $ 3.2 _{ -0.1}^{+ 0.2 }$ &$ 3.9 _{ -0.5}^{+ 0.6 }$ &$ 0.10 _{ -0.05}^{+ 0.05 }$ &$ 13.0 _{ -0.0}^{+ 0.0}$\\
IC 1257 & 1.1 & 1.00 & $ 2.2 _{ -1.1}^{+ 1.1 }$ &$ 18.2 _{ -2.7}^{+ 2.8 }$ &$ 0.79 _{ -0.13}^{+ 0.11 }$ &$ 12.7 _{ -0.5}^{+ 0.3}$\\
& 2.3 & 1.00 & $ 2.1 _{ -1.2}^{+ 1.0 }$ &$ 18.1 _{ -2.6}^{+ 2.8 }$ &$ 0.79 _{ -0.13}^{+ 0.13 }$ &$ 12.5 _{ -0.1}^{+ 0.5}$\\
Ter 2 & 1.1 & 1.00 & $ 0.3 _{ -0.2}^{+ 0.4 }$ &$ 1.4 _{ -0.8}^{+ 19.9 }$ &$ 0.67 _{ -0.60}^{+ 0.32 }$ &$ 12.6 _{ -10.6}^{+ 0.4}$\\
& 2.3 & 1.00 & $ 0.3 _{ -0.2}^{+ 0.3 }$ &$ 1.3 _{ -0.7}^{+ 10.7 }$ &$ 0.64 _{ -0.56}^{+ 0.33 }$ &$ 12.4 _{ -5.9}^{+ 0.6}$\\
NGC 6366 & 1.1 & 1.00 & $ 2.2 _{ -0.1}^{+ 0.1 }$ &$ 5.8 _{ -0.2}^{+ 0.2 }$ &$ 0.44 _{ -0.01}^{+ 0.01 }$ &$ 12.9 _{ -0.1}^{+ 0.1}$\\
& 2.3 & 1.00 & $ 2.2 _{ -0.1}^{+ 0.1 }$ &$ 5.7 _{ -0.2}^{+ 0.2 }$ &$ 0.45 _{ -0.01}^{+ 0.01 }$ &$ 12.9 _{ -0.0}^{+ 0.1}$\\
Ter 4 & 1.1 & 1.00 & $ 0.6 _{ -0.3}^{+ 0.4 }$ &$ 1.7 _{ -0.6}^{+ 0.7 }$ &$ 0.46 _{ -0.10}^{+ 0.13 }$ &$ 13.0 _{ -0.4}^{+ 0.0}$\\
& 2.3 & 1.00 & $ 0.6 _{ -0.3}^{+ 0.3 }$ &$ 1.7 _{ -0.6}^{+ 0.6 }$ &$ 0.47 _{ -0.09}^{+ 0.13 }$ &$ 13.0 _{ -0.4}^{+ 0.0}$\\
HP 1 & 1.1 & 1.00 & $ 0.8 _{ -0.2}^{+ 0.2 }$ &$ 2.2 _{ -0.5}^{+ 0.6 }$ &$ 0.48 _{ -0.14}^{+ 0.11 }$ &$ 13.0 _{ -1.2}^{+ 0.0}$\\
& 2.3 & 1.00 & $ 0.7 _{ -0.1}^{+ 0.3 }$ &$ 2.3 _{ -0.7}^{+ 0.6 }$ &$ 0.52 _{ -0.19}^{+ 0.07 }$ &$ 13.0 _{ -0.7}^{+ 0.0}$\\
NGC 6362 & 1.1 & 1.00 & $ 2.8 _{ -0.1}^{+ 0.1 }$ &$ 5.2 _{ -0.0}^{+ 0.1 }$ &$ 0.30 _{ -0.02}^{+ 0.01 }$ &$ 12.9 _{ -0.1}^{+ 0.1}$\\
& 2.3 & 1.00 & $ 2.8 _{ -0.1}^{+ 0.1 }$ &$ 5.2 _{ -0.0}^{+ 0.1 }$ &$ 0.31 _{ -0.01}^{+ 0.01 }$ &$ 13.0 _{ -0.0}^{+ 0.0}$\\
Lil 1 & 1.1 & 1.00 & $ 0.2 _{ -0.1}^{+ 0.4 }$ &$ 0.9 _{ -0.2}^{+ 8.7 }$ &$ 0.56 _{ -0.18}^{+ 0.40 }$ &$ 13.0 _{ -8.4}^{+ 0.0}$\\
& 2.3 & 1.00 & $ 0.2 _{ -0.1}^{+ 0.3 }$ &$ 0.9 _{ -0.4}^{+ 7.3 }$ &$ 0.56 _{ -0.18}^{+ 0.40 }$ &$ 13.0 _{ -7.0}^{+ 0.0}$\\
NGC 6380 & 1.1 & 1.00 & $ 0.4 _{ -0.2}^{+ 0.3 }$ &$ 2.2 _{ -1.3}^{+ 7.0 }$ &$ 0.69 _{ -0.10}^{+ 0.27 }$ &$ 13.0 _{ -8.2}^{+ 0.0}$\\
& 2.3 & 1.00 & $ 0.4 _{ -0.3}^{+ 0.3 }$ &$ 2.1 _{ -1.4}^{+ 10.5 }$ &$ 0.67 _{ -0.05}^{+ 0.28 }$ &$ 12.7 _{ -5.3}^{+ 0.3}$\\
Ter 1 & 1.1 & 1.00 & $ 0.4 _{ -0.2}^{+ 0.4 }$ &$ 1.8 _{ -1.0}^{+ 3.1 }$ &$ 0.64 _{ -0.12}^{+ 0.29 }$ &$ 12.8 _{ -6.7}^{+ 0.2}$\\
& 2.3 & 1.00 & $ 0.4 _{ -0.2}^{+ 0.4 }$ &$ 1.8 _{ -1.0}^{+ 2.4 }$ &$ 0.65 _{ -0.11}^{+ 0.27 }$ &$ 13.0 _{ -3.8}^{+ 0.0}$\\
Ton 2 & 1.1 & 1.00 & $ 2.2 _{ -0.3}^{+ 0.3 }$ &$ 4.4 _{ -0.7}^{+ 0.8 }$ &$ 0.34 _{ -0.04}^{+ 0.03 }$ &$ 13.0 _{ -0.1}^{+ 0.0}$\\
& 2.3 & 1.00 & $ 2.3 _{ -0.5}^{+ 0.2 }$ &$ 4.1 _{ -0.4}^{+ 0.9 }$ &$ 0.28 _{ -0.01}^{+ 0.09 }$ &$ 13.0 _{ -0.0}^{+ 0.0}$\\
NGC 6388 & 1.1 & 1.00 & $ 1.5 _{ -0.0}^{+ 0.0 }$ &$ 3.5 _{ -0.1}^{+ 0.1 }$ &$ 0.40 _{ -0.01}^{+ 0.01 }$ &$ 13.0 _{ -0.0}^{+ 0.0}$\\
& 2.3 & 1.00 & $ 1.5 _{ -0.0}^{+ 0.0 }$ &$ 3.6 _{ -0.1}^{+ 0.1 }$ &$ 0.42 _{ -0.00}^{+ 0.01 }$ &$ 13.0 _{ 0.0}^{+ 0.0}$\\
NGC 6402 & 1.1 & 1.00 & $ 1.0 _{ -0.4}^{+ 0.2 }$ &$ 4.1 _{ -0.2}^{+ 0.3 }$ &$ 0.61 _{ -0.04}^{+ 0.17 }$ &$ 13.0 _{ -0.4}^{+ 0.0}$\\
& 2.3 & 1.00 & $ 1.0 _{ -0.2}^{+ 0.2 }$ &$ 4.1 _{ -0.1}^{+ 0.2 }$ &$ 0.61 _{ -0.04}^{+ 0.05 }$ &$ 13.0 _{ -0.2}^{+ 0.0}$\\
NGC 6401 & 1.1 & 1.00 & $ 0.4 _{ -0.2}^{+ 0.7 }$ &$ 3.2 _{ -2.1}^{+ 4.8 }$ &$ 0.77 _{ -0.24}^{+ 0.12 }$ &$ 12.8 _{ -8.3}^{+ 0.2}$\\
& 2.3 & 1.00 & $ 0.6 _{ -0.4}^{+ 0.5 }$ &$ 2.8 _{ -1.5}^{+ 10.9 }$ &$ 0.64 _{ -0.11}^{+ 0.31 }$ &$ 13.0 _{ -3.8}^{+ 0.0}$\\
NGC 6397 & 1.1 & 1.00 & $ 2.7 _{ -0.1}^{+ 0.0 }$ &$ 6.4 _{ -0.0}^{+ 0.0 }$ &$ 0.40 _{ -0.00}^{+ 0.01 }$ &$ 12.8 _{ -0.0}^{+ 0.2}$\\
& 2.3 & 1.00 & $ 2.7 _{ -0.1}^{+ 0.0 }$ &$ 6.4 _{ -0.0}^{+ 0.0 }$ &$ 0.41 _{ -0.00}^{+ 0.01 }$ &$ 13.0 _{ -0.1}^{+ 0.0}$\\
Pal 6 & 1.1 & 1.00 & $ 0.9 _{ -0.1}^{+ 0.1 }$ &$ 4.1 _{ -0.6}^{+ 0.6 }$ &$ 0.63 _{ -0.03}^{+ 0.02 }$ &$ 12.5 _{ -0.4}^{+ 0.5}$\\
& 2.3 & 1.00 & $ 0.9 _{ -0.1}^{+ 0.1 }$ &$ 4.0 _{ -0.6}^{+ 0.6 }$ &$ 0.64 _{ -0.02}^{+ 0.02 }$ &$ 12.6 _{ -0.6}^{+ 0.4}$\\
NGC 6426 & 1.1 & 0.60 & $ 27.5 _{ -5.8}^{+ 0.6 }$ &$ 285.3 _{ -198.0}^{+ 46.5 }$ &$ 0.82 _{ -0.22}^{+ 0.02 }$ &$ 5.8 _{ -5.2}^{+ 6.2}$\\
& 2.3 & 1.00 & $ 26.0 _{ -5.0}^{+ 4.7 }$ &$ 100.5 _{ -47.2}^{+ 125.9 }$ &$ 0.59 _{ -0.15}^{+ 0.17 }$ &$ 12.0 _{ -3.4}^{+ 0.2}$\\
Djor 1 & 1.1 & 0.79 & $ 4.2 _{ -1.3}^{+ 5.9 }$ &$ 158.9 _{ -115.6}^{+ 406.3 }$ &$ 0.95 _{ -0.07}^{+ 0.03 }$ &$ 11.1 _{ -10.5}^{+ 1.9}$\\
& 2.3 & 1.00 & $ 4.1 _{ -1.3}^{+ 1.4 }$ &$ 66.9 _{ -36.6}^{+ 107.6 }$ &$ 0.88 _{ -0.06}^{+ 0.05 }$ &$ 11.6 _{ -1.8}^{+ 1.4}$\\
Ter 5 & 1.1 & 1.00 & $ 1.2 _{ -0.3}^{+ 0.4 }$ &$ 3.1 _{ -0.5}^{+ 0.5 }$ &$ 0.46 _{ -0.05}^{+ 0.06 }$ &$ 13.0 _{ -0.0}^{+ 0.0}$\\
& 2.3 & 1.00 & $ 1.1 _{ -0.3}^{+ 0.4 }$ &$ 3.1 _{ -0.5}^{+ 0.5 }$ &$ 0.46 _{ -0.05}^{+ 0.07 }$ &$ 13.0 _{ 0.0}^{+ 0.0}$\\
NGC 6440 & 1.1 & 1.00 & $ 0.3 _{ -0.1}^{+ 0.4 }$ &$ 1.0 _{ -0.2}^{+ 5.9 }$ &$ 0.59 _{ -0.08}^{+ 0.33 }$ &$ 12.0 _{ -6.7}^{+ 1.0}$\\
& 2.3 & 1.00 & $ 0.4 _{ -0.2}^{+ 0.3 }$ &$ 2.8 _{ -2.2}^{+ 4.6 }$ &$ 0.78 _{ -0.26}^{+ 0.15 }$ &$ 12.3 _{ -8.0}^{+ 0.7}$\\
NGC 6441 & 1.1 & 1.00 & $ 1.2 _{ -0.1}^{+ 0.1 }$ &$ 3.5 _{ -0.1}^{+ 0.3 }$ &$ 0.48 _{ -0.02}^{+ 0.06 }$ &$ 13.0 _{ -0.1}^{+ 0.0}$\\
& 2.3 & 1.00 & $ 1.3 _{ -0.1}^{+ 0.1 }$ &$ 3.6 _{ -0.1}^{+ 0.1 }$ &$ 0.48 _{ -0.02}^{+ 0.01 }$ &$ 13.0 _{ -0.0}^{+ 0.0}$\\
Ter 6 & 1.1 & 1.00 & $ 0.3 _{ -0.1}^{+ 0.3 }$ &$ 2.1 _{ -0.8}^{+ 8.6 }$ &$ 0.77 _{ -0.23}^{+ 0.19 }$ &$ 12.7 _{ -7.4}^{+ 0.3}$\\
& 2.3 & 1.00 & $ 0.3 _{ -0.1}^{+ 0.2 }$ &$ 2.2 _{ -0.9}^{+ 6.9 }$ &$ 0.78 _{ -0.25}^{+ 0.18 }$ &$ 12.3 _{ -6.7}^{+ 0.7}$\\
NGC 6453 & 1.1 & 1.00 & $ 1.1 _{ -0.5}^{+ 0.9 }$ &$ 3.9 _{ -1.7}^{+ 1.5 }$ &$ 0.56 _{ -0.10}^{+ 0.08 }$ &$ 13.0 _{ -1.7}^{+ 0.0}$\\
& 2.3 & 1.00 & $ 1.2 _{ -0.5}^{+ 0.6 }$ &$ 3.8 _{ -1.4}^{+ 1.7 }$ &$ 0.53 _{ -0.03}^{+ 0.11 }$ &$ 13.0 _{ -3.5}^{+ 0.0}$\\
UKS 1 & 1.1 & 1.00 & $ 0.6 _{ -0.4}^{+ 0.7 }$ &$ 0.9 _{ -0.1}^{+ 2.7 }$ &$ 0.20 _{ -0.16}^{+ 0.70 }$ &$ 13.0 _{ -3.3}^{+ 0.0}$\\
& 2.3 & 1.00 & $ 0.6 _{ -0.4}^{+ 0.7 }$ &$ 0.9 _{ -0.1}^{+ 2.1 }$ &$ 0.20 _{ -0.16}^{+ 0.65 }$ &$ 13.0 _{ -2.0}^{+ 0.0}$\\
NGC 6496 & 1.1 & 1.00 & $ 3.9 _{ -1.0}^{+ 1.0 }$ &$ 10.8 _{ -4.3}^{+ 9.3 }$ &$ 0.47 _{ -0.09}^{+ 0.14 }$ &$ 12.6 _{ -0.3}^{+ 0.4}$\\
& 2.3 & 1.00 & $ 3.9 _{ -1.0}^{+ 1.0 }$ &$ 9.8 _{ -3.7}^{+ 6.9 }$ &$ 0.43 _{ -0.08}^{+ 0.11 }$ &$ 12.7 _{ -0.2}^{+ 0.3}$\\
Ter 9 & 1.1 & 1.00 & $ 0.3 _{ -0.1}^{+ 0.4 }$ &$ 4.3 _{ -3.7}^{+ 3.6 }$ &$ 0.88 _{ -0.39}^{+ 0.08 }$ &$ 12.7 _{ -6.3}^{+ 0.3}$\\
& 2.3 & 1.00 & $ 0.3 _{ -0.1}^{+ 0.5 }$ &$ 1.2 _{ -0.7}^{+ 8.6 }$ &$ 0.65 _{ -0.15}^{+ 0.31 }$ &$ 12.7 _{ -5.4}^{+ 0.3}$\\
Djor 2 & 1.1 & 1.00 & $ 1.1 _{ -0.4}^{+ 0.4 }$ &$ 3.4 _{ -0.7}^{+ 0.7 }$ &$ 0.52 _{ -0.05}^{+ 0.09 }$ &$ 13.0 _{ -0.1}^{+ 0.0}$\\
& 2.3 & 1.00 & $ 1.1 _{ -0.4}^{+ 0.4 }$ &$ 3.4 _{ -0.7}^{+ 0.7 }$ &$ 0.52 _{ -0.06}^{+ 0.08 }$ &$ 13.0 _{ 0.0}^{+ 0.0}$\\
NGC 6517 & 1.1 & 1.00 & $ 0.8 _{ -0.1}^{+ 0.2 }$ &$ 4.0 _{ -0.6}^{+ 0.8 }$ &$ 0.65 _{ -0.02}^{+ 0.01 }$ &$ 12.9 _{ -0.6}^{+ 0.1}$\\
& 2.3 & 1.00 & $ 0.8 _{ -0.1}^{+ 0.2 }$ &$ 4.0 _{ -0.6}^{+ 0.8 }$ &$ 0.66 _{ -0.02}^{+ 0.01 }$ &$ 13.0 _{ -0.4}^{+ 0.0}$\\
Ter 10 & 1.1 & 1.00 & $ 1.1 _{ -0.5}^{+ 0.3 }$ &$ 2.6 _{ -0.5}^{+ 0.7 }$ &$ 0.40 _{ -0.02}^{+ 0.21 }$ &$ 13.0 _{ -0.1}^{+ 0.0}$\\
& 2.3 & 1.00 & $ 1.1 _{ -0.5}^{+ 0.3 }$ &$ 2.7 _{ -0.5}^{+ 0.7 }$ &$ 0.41 _{ -0.03}^{+ 0.17 }$ &$ 13.0 _{ -0.0}^{+ 0.0}$\\
NGC 6522 & 1.1 & 1.00 & $ 0.4 _{ -0.3}^{+ 0.6 }$ &$ 1.2 _{ -0.6}^{+ 10.7 }$ &$ 0.50 _{ -0.27}^{+ 0.37 }$ &$ 12.7 _{ -8.9}^{+ 0.3}$\\
& 2.3 & 1.00 & $ 0.4 _{ -0.3}^{+ 0.5 }$ &$ 1.2 _{ -0.6}^{+ 8.0 }$ &$ 0.47 _{ -0.24}^{+ 0.39 }$ &$ 12.9 _{ -6.0}^{+ 0.1}$\\
NGC 6535 & 1.1 & 1.00 & $ 1.3 _{ -0.5}^{+ 0.4 }$ &$ 4.4 _{ -0.2}^{+ 0.7 }$ &$ 0.55 _{ -0.11}^{+ 0.19 }$ &$ 13.0 _{ -0.3}^{+ 0.0}$\\
& 2.3 & 1.00 & $ 1.3 _{ -0.6}^{+ 0.3 }$ &$ 4.4 _{ -0.2}^{+ 0.5 }$ &$ 0.55 _{ -0.10}^{+ 0.19 }$ &$ 13.0 _{ -0.1}^{+ 0.0}$\\
NGC 6528 & 1.1 & 1.00 & $ 0.6 _{ -0.3}^{+ 0.1 }$ &$ 2.3 _{ -0.5}^{+ 1.1 }$ &$ 0.57 _{ -0.01}^{+ 0.16 }$ &$ 13.0 _{ -2.4}^{+ 0.0}$\\
& 2.3 & 1.00 & $ 0.6 _{ -0.3}^{+ 0.1 }$ &$ 2.2 _{ -0.7}^{+ 1.0 }$ &$ 0.57 _{ -0.00}^{+ 0.22 }$ &$ 13.0 _{ -2.8}^{+ 0.0}$\\
NGC 6539 & 1.1 & 1.00 & $ 2.3 _{ -0.2}^{+ 0.2 }$ &$ 3.3 _{ -0.1}^{+ 0.2 }$ &$ 0.19 _{ -0.02}^{+ 0.03 }$ &$ 13.0 _{ -0.0}^{+ 0.0}$\\
& 2.3 & 1.00 & $ 2.2 _{ -0.1}^{+ 0.2 }$ &$ 3.3 _{ -0.1}^{+ 0.2 }$ &$ 0.21 _{ -0.03}^{+ 0.02 }$ &$ 13.0 _{ 0.0}^{+ 0.0}$\\
NGC 6540 & 1.1 & 1.00 & $ 1.9 _{ -0.4}^{+ 0.4 }$ &$ 3.2 _{ -0.5}^{+ 0.5 }$ &$ 0.24 _{ -0.02}^{+ 0.02 }$ &$ 13.0 _{ -0.0}^{+ 0.0}$\\
& 2.3 & 1.00 & $ 1.9 _{ -0.4}^{+ 0.4 }$ &$ 3.2 _{ -0.5}^{+ 0.5 }$ &$ 0.26 _{ -0.02}^{+ 0.02 }$ &$ 13.0 _{ 0.0}^{+ 0.0}$\\
NGC 6544 & 1.1 & 1.00 & $ 0.6 _{ -0.3}^{+ 0.4 }$ &$ 5.8 _{ -0.5}^{+ 0.2 }$ &$ 0.82 _{ -0.09}^{+ 0.08 }$ &$ 12.9 _{ -3.4}^{+ 0.1}$\\
& 2.3 & 1.00 & $ 0.6 _{ -0.3}^{+ 0.4 }$ &$ 5.8 _{ -2.5}^{+ 5.1 }$ &$ 0.83 _{ -0.09}^{+ 0.12 }$ &$ 12.9 _{ -3.3}^{+ 0.1}$\\
NGC 6540 & 1.1 & 1.00 & $ 1.8 _{ -0.1}^{+ 0.1 }$ &$ 4.4 _{ -0.1}^{+ 0.2 }$ &$ 0.43 _{ -0.02}^{+ 0.02 }$ &$ 12.9 _{ -0.0}^{+ 0.1}$\\
& 2.3 & 1.00 & $ 1.8 _{ -0.1}^{+ 0.1 }$ &$ 4.3 _{ -0.1}^{+ 0.2 }$ &$ 0.41 _{ -0.01}^{+ 0.03 }$ &$ 13.0 _{ -0.0}^{+ 0.0}$\\
ESO 280-SC06 & 1.1 & 1.00 & $ 2.9 _{ -1.6}^{+ 2.9 }$ &$ 16.1 _{ -2.9}^{+ 2.7 }$ &$ 0.69 _{ -0.17}^{+ 0.13 }$ &$ 12.7 _{ -0.4}^{+ 0.3}$\\
& 2.3 & 1.00 & $ 2.8 _{ -1.6}^{+ 2.6 }$ &$ 15.9 _{ -2.8}^{+ 2.6 }$ &$ 0.70 _{ -0.15}^{+ 0.13 }$ &$ 12.6 _{ -0.2}^{+ 0.4}$\\
NGC 6553 & 1.1 & 1.00 & $ 1.6 _{ -0.2}^{+ 0.2 }$ &$ 3.2 _{ -0.2}^{+ 0.1 }$ &$ 0.34 _{ -0.05}^{+ 0.05 }$ &$ 13.0 _{ 0.0}^{+ 0.0}$\\
& 2.3 & 1.00 & $ 1.6 _{ -0.2}^{+ 0.2 }$ &$ 3.1 _{ -0.2}^{+ 0.2 }$ &$ 0.32 _{ -0.04}^{+ 0.05 }$ &$ 13.0 _{ 0.0}^{+ 0.0}$\\
2MS 2 & 1.1 & 1.00 & $ 0.4 _{ -0.3}^{+ 0.5 }$ &$ 3.1 _{ -1.6}^{+ 9.5 }$ &$ 0.75 _{ -0.30}^{+ 0.22 }$ &$ 13.0 _{ -8.0}^{+ 0.0}$\\
& 2.3 & 1.00 & $ 0.4 _{ -0.3}^{+ 0.5 }$ &$ 3.1 _{ -1.2}^{+ 9.7 }$ &$ 0.75 _{ -0.30}^{+ 0.23 }$ &$ 13.0 _{ -5.3}^{+ 0.0}$\\
NGC 6558 & 1.1 & 1.00 & $ 0.7 _{ -0.5}^{+ 0.1 }$ &$ 2.5 _{ -1.1}^{+ 2.6 }$ &$ 0.57 _{ -0.02}^{+ 0.32 }$ &$ 13.0 _{ -5.2}^{+ 0.0}$\\
& 2.3 & 1.00 & $ 0.7 _{ -0.4}^{+ 0.1 }$ &$ 2.5 _{ -1.0}^{+ 1.5 }$ &$ 0.58 _{ -0.03}^{+ 0.24 }$ &$ 13.0 _{ -3.1}^{+ 0.0}$\\
IC 1276 & 1.1 & 1.00 & $ 3.6 _{ -0.1}^{+ 0.2 }$ &$ 6.2 _{ -0.6}^{+ 0.7 }$ &$ 0.27 _{ -0.03}^{+ 0.03 }$ &$ 12.9 _{ -0.1}^{+ 0.1}$\\
& 2.3 & 1.00 & $ 3.6 _{ -0.1}^{+ 0.2 }$ &$ 6.0 _{ -0.6}^{+ 0.7 }$ &$ 0.25 _{ -0.03}^{+ 0.03 }$ &$ 12.9 _{ -0.0}^{+ 0.1}$\\
Ter 12 & 1.1 & 1.00 & $ 3.2 _{ -0.3}^{+ 0.3 }$ &$ 6.2 _{ -0.5}^{+ 0.5 }$ &$ 0.32 _{ -0.01}^{+ 0.01 }$ &$ 13.0 _{ -0.2}^{+ 0.0}$\\
& 2.3 & 1.00 & $ 3.2 _{ -0.3}^{+ 0.3 }$ &$ 6.1 _{ -0.5}^{+ 0.5 }$ &$ 0.32 _{ -0.01}^{+ 0.01 }$ &$ 12.9 _{ -0.0}^{+ 0.1}$\\
NGC 6569 & 1.1 & 1.00 & $ 2.1 _{ -0.7}^{+ 0.8 }$ &$ 2.7 _{ -0.9}^{+ 1.1 }$ &$ 0.13 _{ -0.01}^{+ 0.03 }$ &$ 13.0 _{ -0.0}^{+ 0.0}$\\
& 2.3 & 1.00 & $ 2.0 _{ -0.7}^{+ 0.8 }$ &$ 2.7 _{ -0.9}^{+ 1.1 }$ &$ 0.13 _{ -0.01}^{+ 0.02 }$ &$ 13.0 _{ 0.0}^{+ 0.0}$\\
BH 261 & 1.1 & 1.00 & $ 1.7 _{ -0.5}^{+ 0.6 }$ &$ 3.2 _{ -0.5}^{+ 0.4 }$ &$ 0.32 _{ -0.10}^{+ 0.08 }$ &$ 13.0 _{ -0.0}^{+ 0.0}$\\
& 2.3 & 1.00 & $ 1.7 _{ -0.5}^{+ 0.6 }$ &$ 3.1 _{ -0.5}^{+ 0.4 }$ &$ 0.29 _{ -0.08}^{+ 0.11 }$ &$ 13.0 _{ 0.0}^{+ 0.0}$\\
NGC 6584 & 1.1 & 1.00 & $ 2.0 _{ -0.8}^{+ 1.4 }$ &$ 19.6 _{ -6.5}^{+ 11.6 }$ &$ 0.81 _{ -0.01}^{+ 0.02 }$ &$ 12.4 _{ -0.5}^{+ 0.6}$\\
& 2.3 & 1.00 & $ 2.0 _{ -0.9}^{+ 1.3 }$ &$ 17.0 _{ -5.0}^{+ 7.6 }$ &$ 0.79 _{ -0.03}^{+ 0.04 }$ &$ 12.7 _{ -0.8}^{+ 0.3}$\\
Mercer 5 & 1.1 & 1.00 & $ 2.2 _{ -0.2}^{+ 0.3 }$ &$ 6.1 _{ -0.8}^{+ 0.9 }$ &$ 0.47 _{ -0.06}^{+ 0.04 }$ &$ 12.8 _{ -0.1}^{+ 0.2}$\\
& 2.3 & 1.00 & $ 2.2 _{ -0.2}^{+ 0.3 }$ &$ 5.9 _{ -0.7}^{+ 0.8 }$ &$ 0.46 _{ -0.06}^{+ 0.04 }$ &$ 13.0 _{ -0.1}^{+ 0.0}$\\
NGC 6624 & 1.1 & 1.00 & $ 0.9 _{ -0.1}^{+ 0.0 }$ &$ 1.5 _{ -0.1}^{+ 0.3 }$ &$ 0.23 _{ -0.00}^{+ 0.13 }$ &$ 13.0 _{ -0.4}^{+ 0.0}$\\
& 2.3 & 1.00 & $ 0.9 _{ -0.1}^{+ 0.0 }$ &$ 1.5 _{ -0.1}^{+ 0.4 }$ &$ 0.23 _{ -0.00}^{+ 0.16 }$ &$ 13.0 _{ -0.4}^{+ 0.0}$\\
NGC 6626 & 1.1 & 1.00 & $ 0.7 _{ -0.2}^{+ 0.1 }$ &$ 3.1 _{ -0.4}^{+ 0.3 }$ &$ 0.64 _{ -0.06}^{+ 0.07 }$ &$ 13.0 _{ -0.2}^{+ 0.0}$\\
& 2.3 & 1.00 & $ 0.7 _{ -0.2}^{+ 0.1 }$ &$ 3.0 _{ -0.3}^{+ 0.5 }$ &$ 0.61 _{ -0.02}^{+ 0.12 }$ &$ 13.0 _{ -0.2}^{+ 0.0}$\\
NGC 6638 & 1.1 & 1.00 & $ 0.5 _{ -0.3}^{+ 0.3 }$ &$ 3.1 _{ -2.0}^{+ 6.0 }$ &$ 0.72 _{ -0.18}^{+ 0.21 }$ &$ 12.4 _{ -6.4}^{+ 0.6}$\\
& 2.3 & 1.00 & $ 0.3 _{ -0.1}^{+ 0.6 }$ &$ 4.5 _{ -3.5}^{+ 5.1 }$ &$ 0.88 _{ -0.35}^{+ 0.08 }$ &$ 12.9 _{ -3.8}^{+ 0.1}$\\
NGC 6637 & 1.1 & 1.00 & $ 1.0 _{ -0.7}^{+ 0.5 }$ &$ 1.7 _{ -0.8}^{+ 3.0 }$ &$ 0.26 _{ -0.07}^{+ 0.44 }$ &$ 13.0 _{ -6.0}^{+ 0.0}$\\
& 2.3 & 1.00 & $ 1.0 _{ -0.7}^{+ 0.5 }$ &$ 1.7 _{ -0.8}^{+ 4.1 }$ &$ 0.26 _{ -0.07}^{+ 0.47 }$ &$ 13.0 _{ -3.6}^{+ 0.0}$\\
NGC 6642 & 1.1 & 1.00 & $ 0.2 _{ -0.1}^{+ 0.6 }$ &$ 1.7 _{ -1.0}^{+ 9.1 }$ &$ 0.77 _{ -0.30}^{+ 0.19 }$ &$ 12.6 _{ -7.0}^{+ 0.4}$\\
& 2.3 & 1.00 & $ 0.3 _{ -0.2}^{+ 0.5 }$ &$ 2.2 _{ -1.1}^{+ 10.8 }$ &$ 0.72 _{ -0.25}^{+ 0.24 }$ &$ 12.6 _{ -5.7}^{+ 0.4}$\\
NGC 6652 & 1.1 & 1.00 & $ 0.8 _{ -0.6}^{+ 0.7 }$ &$ 3.7 _{ -2.9}^{+ 4.6 }$ &$ 0.66 _{ -0.14}^{+ 0.16 }$ &$ 12.8 _{ -7.2}^{+ 0.2}$\\
& 2.3 & 1.00 & $ 0.8 _{ -0.5}^{+ 0.7 }$ &$ 3.6 _{ -2.5}^{+ 6.0 }$ &$ 0.65 _{ -0.14}^{+ 0.16 }$ &$ 12.1 _{ -4.0}^{+ 0.9}$\\
NGC 6656 & 1.1 & 1.00 & $ 3.2 _{ -0.0}^{+ 0.1 }$ &$ 10.1 _{ -0.0}^{+ 0.1 }$ &$ 0.52 _{ -0.01}^{+ 0.00 }$ &$ 12.7 _{ -0.1}^{+ 0.3}$\\
& 2.3 & 1.00 & $ 3.2 _{ -0.1}^{+ 0.0 }$ &$ 9.5 _{ -0.0}^{+ 0.0 }$ &$ 0.50 _{ -0.01}^{+ 0.01 }$ &$ 12.9 _{ -0.1}^{+ 0.1}$\\
Pal 8 & 1.1 & 1.00 & $ 2.3 _{ -0.7}^{+ 1.4 }$ &$ 5.3 _{ -1.2}^{+ 1.3 }$ &$ 0.40 _{ -0.12}^{+ 0.05 }$ &$ 13.0 _{ -0.2}^{+ 0.0}$\\
& 2.3 & 1.00 & $ 2.2 _{ -0.7}^{+ 1.3 }$ &$ 5.2 _{ -1.1}^{+ 1.3 }$ &$ 0.41 _{ -0.11}^{+ 0.06 }$ &$ 13.0 _{ -0.1}^{+ 0.0}$\\
NGC 6681 & 1.1 & 1.00 & $ 0.8 _{ -0.5}^{+ 0.4 }$ &$ 3.8 _{ -2.2}^{+ 13.3 }$ &$ 0.66 _{ -0.05}^{+ 0.22 }$ &$ 13.0 _{ -8.3}^{+ 0.0}$\\
& 2.3 & 1.00 & $ 1.0 _{ -0.6}^{+ 0.2 }$ &$ 5.8 _{ -3.9}^{+ 8.0 }$ &$ 0.70 _{ -0.09}^{+ 0.15 }$ &$ 12.6 _{ -3.3}^{+ 0.4}$\\
NGC 6712 & 1.1 & 1.00 & $ 1.0 _{ -0.1}^{+ 0.1 }$ &$ 4.9 _{ -0.0}^{+ 0.1 }$ &$ 0.66 _{ -0.02}^{+ 0.02 }$ &$ 12.6 _{ -0.7}^{+ 0.4}$\\
& 2.3 & 1.00 & $ 1.0 _{ -0.1}^{+ 0.1 }$ &$ 4.9 _{ -0.0}^{+ 0.1 }$ &$ 0.67 _{ -0.02}^{+ 0.02 }$ &$ 13.0 _{ -1.1}^{+ 0.0}$\\
NGC 6715 & 1.1 & 1.00 & $ 12.7 _{ -0.6}^{+ 0.5 }$ &$ 41.2 _{ -3.2}^{+ 3.7 }$ &$ 0.53 _{ -0.01}^{+ 0.02 }$ &$ 11.5 _{ -0.1}^{+ 1.0}$\\
& 2.3 & 1.00 & $ 12.2 _{ -0.6}^{+ 0.6 }$ &$ 31.0 _{ -1.8}^{+ 1.9 }$ &$ 0.44 _{ -0.01}^{+ 0.01 }$ &$ 12.2 _{ -0.2}^{+ 0.5}$\\
NGC 6717 & 1.1 & 1.00 & $ 1.2 _{ -0.0}^{+ 0.5 }$ &$ 2.8 _{ -0.4}^{+ 0.2 }$ &$ 0.40 _{ -0.15}^{+ 0.00 }$ &$ 13.0 _{ 0.0}^{+ 0.0}$\\
& 2.3 & 1.00 & $ 1.3 _{ -0.0}^{+ 0.4 }$ &$ 2.6 _{ -0.2}^{+ 0.4 }$ &$ 0.35 _{ -0.07}^{+ 0.01 }$ &$ 13.0 _{ 0.0}^{+ 0.0}$\\
NGC 6723 & 1.1 & 1.00 & $ 1.5 _{ -0.0}^{+ 0.2 }$ &$ 3.8 _{ -0.1}^{+ 0.1 }$ &$ 0.45 _{ -0.06}^{+ 0.00 }$ &$ 13.0 _{ -0.1}^{+ 0.0}$\\
& 2.3 & 1.00 & $ 1.5 _{ -0.0}^{+ 0.2 }$ &$ 3.7 _{ -0.1}^{+ 0.1 }$ &$ 0.41 _{ -0.05}^{+ 0.01 }$ &$ 13.0 _{ -0.0}^{+ 0.0}$\\
NGC 6749 & 1.1 & 1.00 & $ 1.7 _{ -0.2}^{+ 0.3 }$ &$ 5.1 _{ -0.1}^{+ 0.3 }$ &$ 0.50 _{ -0.05}^{+ 0.03 }$ &$ 12.9 _{ -0.0}^{+ 0.1}$\\
& 2.3 & 1.00 & $ 1.6 _{ -0.2}^{+ 0.3 }$ &$ 5.1 _{ -0.1}^{+ 0.3 }$ &$ 0.51 _{ -0.05}^{+ 0.03 }$ &$ 13.0 _{ -0.0}^{+ 0.0}$\\
NGC 6752 & 1.1 & 1.00 & $ 3.7 _{ -0.2}^{+ 0.1 }$ &$ 5.6 _{ -0.0}^{+ 0.0 }$ &$ 0.20 _{ -0.02}^{+ 0.02 }$ &$ 12.8 _{ -0.0}^{+ 0.1}$\\
& 2.3 & 1.00 & $ 3.5 _{ -0.1}^{+ 0.0 }$ &$ 5.5 _{ -0.1}^{+ 0.1 }$ &$ 0.23 _{ -0.00}^{+ 0.01 }$ &$ 13.0 _{ -0.0}^{+ 0.0}$\\
NGC 6760 & 1.1 & 1.00 & $ 2.2 _{ -0.1}^{+ 0.2 }$ &$ 5.7 _{ -0.2}^{+ 0.2 }$ &$ 0.45 _{ -0.04}^{+ 0.04 }$ &$ 13.0 _{ -0.2}^{+ 0.0}$\\
& 2.3 & 1.00 & $ 2.1 _{ -0.1}^{+ 0.1 }$ &$ 5.6 _{ -0.1}^{+ 0.2 }$ &$ 0.45 _{ -0.03}^{+ 0.04 }$ &$ 12.9 _{ -0.0}^{+ 0.1}$\\
NGC 6779 & 1.1 & 1.00 & $ 1.1 _{ -0.0}^{+ 0.3 }$ &$ 12.8 _{ -1.0}^{+ 0.9 }$ &$ 0.85 _{ -0.02}^{+ 0.01 }$ &$ 12.5 _{ -0.8}^{+ 0.5}$\\
& 2.3 & 1.00 & $ 0.9 _{ -0.0}^{+ 0.4 }$ &$ 12.4 _{ -0.9}^{+ 0.7 }$ &$ 0.86 _{ -0.04}^{+ 0.00 }$ &$ 12.7 _{ -0.7}^{+ 0.3}$\\
Ter 7 & 1.1 & 1.00 & $ 13.1 _{ -2.4}^{+ 2.5 }$ &$ 47.2 _{ -15.6}^{+ 27.1 }$ &$ 0.57 _{ -0.07}^{+ 0.09 }$ &$ 11.4 _{ -0.5}^{+ 1.1}$\\
& 2.3 & 1.00 & $ 12.8 _{ -2.4}^{+ 2.5 }$ &$ 33.5 _{ -8.7}^{+ 12.5 }$ &$ 0.45 _{ -0.04}^{+ 0.05 }$ &$ 12.1 _{ -0.4}^{+ 0.6}$\\
Pal 10 & 1.1 & 1.00 & $ 4.0 _{ -0.2}^{+ 0.3 }$ &$ 7.2 _{ -0.2}^{+ 0.3 }$ &$ 0.28 _{ -0.02}^{+ 0.01 }$ &$ 12.7 _{ -0.0}^{+ 0.2}$\\
& 2.3 & 1.00 & $ 3.9 _{ -0.2}^{+ 0.3 }$ &$ 7.1 _{ -0.2}^{+ 0.3 }$ &$ 0.29 _{ -0.02}^{+ 0.02 }$ &$ 12.9 _{ -0.1}^{+ 0.1}$\\
Arp 2 & 1.1 & 1.00 & $ 18.2 _{ -3.4}^{+ 3.4 }$ &$ 65.2 _{ -23.1}^{+ 45.2 }$ &$ 0.56 _{ -0.08}^{+ 0.11 }$ &$ 12.0 _{ -2.3}^{+ 0.3}$\\
& 2.3 & 1.00 & $ 17.7 _{ -3.5}^{+ 3.5 }$ &$ 43.1 _{ -11.4}^{+ 17.5 }$ &$ 0.42 _{ -0.04}^{+ 0.07 }$ &$ 11.9 _{ -0.7}^{+ 0.6}$\\
NGC 6809 & 1.1 & 1.00 & $ 1.6 _{ -0.0}^{+ 0.2 }$ &$ 6.2 _{ -0.3}^{+ 0.2 }$ &$ 0.59 _{ -0.05}^{+ 0.01 }$ &$ 12.9 _{ -0.2}^{+ 0.1}$\\
& 2.3 & 1.00 & $ 1.6 _{ -0.1}^{+ 0.0 }$ &$ 6.0 _{ -0.2}^{+ 0.2 }$ &$ 0.58 _{ -0.02}^{+ 0.01 }$ &$ 13.0 _{ -0.2}^{+ 0.0}$\\
Ter 8 & 1.1 & 1.00 & $ 16.8 _{ -2.9}^{+ 3.0 }$ &$ 62.9 _{ -20.7}^{+ 37.5 }$ &$ 0.58 _{ -0.07}^{+ 0.09 }$ &$ 11.1 _{ -1.0}^{+ 1.2}$\\
& 2.3 & 1.00 & $ 16.4 _{ -3.0}^{+ 3.0 }$ &$ 41.8 _{ -10.4}^{+ 15.2 }$ &$ 0.44 _{ -0.04}^{+ 0.06 }$ &$ 12.0 _{ -0.7}^{+ 0.5}$\\
Pal 11 & 1.1 & 1.00 & $ 5.2 _{ -1.8}^{+ 2.9 }$ &$ 8.9 _{ -1.2}^{+ 1.4 }$ &$ 0.26 _{ -0.14}^{+ 0.12 }$ &$ 12.7 _{ -0.1}^{+ 0.3}$\\
& 2.3 & 1.00 & $ 4.9 _{ -1.6}^{+ 2.5 }$ &$ 8.9 _{ -1.2}^{+ 1.4 }$ &$ 0.29 _{ -0.13}^{+ 0.12 }$ &$ 12.8 _{ -0.1}^{+ 0.2}$\\
NGC 6838 & 1.1 & 1.00 & $ 5.0 _{ -0.1}^{+ 0.1 }$ &$ 7.3 _{ -0.0}^{+ 0.0 }$ &$ 0.18 _{ -0.01}^{+ 0.01 }$ &$ 12.8 _{ -0.1}^{+ 0.0}$\\
& 2.3 & 1.00 & $ 4.8 _{ -0.1}^{+ 0.1 }$ &$ 7.2 _{ -0.0}^{+ 0.0 }$ &$ 0.20 _{ -0.01}^{+ 0.01 }$ &$ 12.9 _{ -0.1}^{+ 0.1}$\\
NGC 6864 & 1.1 & 1.00 & $ 1.9 _{ -0.8}^{+ 1.3 }$ &$ 17.6 _{ -3.5}^{+ 3.2 }$ &$ 0.80 _{ -0.08}^{+ 0.07 }$ &$ 12.7 _{ -5.3}^{+ 0.3}$\\
& 2.3 & 1.00 & $ 1.8 _{ -0.7}^{+ 1.3 }$ &$ 17.2 _{ -4.0}^{+ 2.8 }$ &$ 0.81 _{ -0.08}^{+ 0.06 }$ &$ 12.6 _{ -3.1}^{+ 0.4}$\\
NGC 6934 & 1.1 & 1.00 & $ 2.6 _{ -0.9}^{+ 1.2 }$ &$ 45.1 _{ -10.0}^{+ 14.9 }$ &$ 0.89 _{ -0.01}^{+ 0.02 }$ &$ 11.9 _{ -0.5}^{+ 1.1}$\\
& 2.3 & 1.00 & $ 2.6 _{ -0.9}^{+ 1.1 }$ &$ 34.1 _{ -5.8}^{+ 7.4 }$ &$ 0.86 _{ -0.02}^{+ 0.03 }$ &$ 13.0 _{ -1.0}^{+ 0.0}$\\
NGC 6981 & 1.1 & 1.00 & $ 1.2 _{ -0.0}^{+ 0.7 }$ &$ 24.6 _{ -4.5}^{+ 5.4 }$ &$ 0.91 _{ -0.07}^{+ 0.00 }$ &$ 11.0 _{ -1.8}^{+ 2.0}$\\
& 2.3 & 1.00 & $ 1.2 _{ -0.3}^{+ 0.7 }$ &$ 24.1 _{ -8.8}^{+ 10.5 }$ &$ 0.90 _{ -0.07}^{+ 0.04 }$ &$ 12.6 _{ -4.4}^{+ 0.4}$\\
NGC 7006 & 1.1 & 1.00 & $ 2.2 _{ -1.7}^{+ 1.7 }$ &$ 58.6 _{ -7.1}^{+ 9.7 }$ &$ 0.93 _{ -0.07}^{+ 0.06 }$ &$ 12.5 _{ -3.0}^{+ 0.5}$\\
& 2.3 & 1.00 & $ 2.0 _{ -1.3}^{+ 1.7 }$ &$ 52.0 _{ -5.9}^{+ 14.1 }$ &$ 0.93 _{ -0.07}^{+ 0.05 }$ &$ 11.8 _{ -1.0}^{+ 1.2}$\\
NGC 7078 & 1.1 & 1.00 & $ 3.9 _{ -0.1}^{+ 0.1 }$ &$ 10.5 _{ -0.1}^{+ 0.1 }$ &$ 0.45 _{ -0.01}^{+ 0.01 }$ &$ 12.6 _{ -0.0}^{+ 0.4}$\\
& 2.3 & 1.00 & $ 3.8 _{ -0.1}^{+ 0.1 }$ &$ 10.5 _{ -0.1}^{+ 0.1 }$ &$ 0.47 _{ -0.01}^{+ 0.01 }$ &$ 12.9 _{ -0.2}^{+ 0.1}$\\
NGC 7089 & 1.1 & 1.00 & $ 1.2 _{ -0.0}^{+ 0.1 }$ &$ 17.8 _{ -1.0}^{+ 0.9 }$ &$ 0.88 _{ -0.02}^{+ 0.01 }$ &$ 12.6 _{ -1.9}^{+ 0.4}$\\
& 2.3 & 1.00 & $ 1.1 _{ -0.0}^{+ 0.1 }$ &$ 16.4 _{ -0.6}^{+ 0.8 }$ &$ 0.87 _{ -0.01}^{+ 0.01 }$ &$ 12.0 _{ -0.9}^{+ 1.0}$\\
NGC 7099 & 1.1 & 1.00 & $ 2.0 _{ -0.1}^{+ 0.0 }$ &$ 8.2 _{ -0.6}^{+ 0.7 }$ &$ 0.61 _{ -0.02}^{+ 0.02 }$ &$ 12.9 _{ -0.3}^{+ 0.1}$\\
& 2.3 & 1.00 & $ 2.0 _{ -0.1}^{+ 0.0 }$ &$ 8.1 _{ -0.5}^{+ 0.6 }$ &$ 0.61 _{ -0.01}^{+ 0.02 }$ &$ 12.9 _{ -0.1}^{+ 0.1}$\\
Pal 12 & 1.1 & 1.00 & $ 15.6 _{ -1.8}^{+ 2.0 }$ &$ 79.1 _{ -32.8}^{+ 76.7 }$ &$ 0.67 _{ -0.13}^{+ 0.13 }$ &$ 10.9 _{ -1.0}^{+ 1.3}$\\
& 2.3 & 1.00 & $ 15.6 _{ -1.8}^{+ 1.8 }$ &$ 46.7 _{ -14.6}^{+ 23.5 }$ &$ 0.50 _{ -0.10}^{+ 0.10 }$ &$ 12.4 _{ -1.3}^{+ 0.1}$\\
Pal 13 & 1.1 & 1.00 & $ 8.7 _{ -2.0}^{+ 2.8 }$ &$ 78.4 _{ -17.0}^{+ 27.4 }$ &$ 0.80 _{ -0.01}^{+ 0.01 }$ &$ 11.7 _{ -1.6}^{+ 1.1}$\\
& 2.3 & 1.00 & $ 8.2 _{ -1.9}^{+ 2.6 }$ &$ 54.7 _{ -8.5}^{+ 11.5 }$ &$ 0.74 _{ -0.02}^{+ 0.03 }$ &$ 12.4 _{ -1.1}^{+ 0.4}$\\
NGC 7492 & 1.1 & 1.00 & $ 4.6 _{ -2.2}^{+ 3.7 }$ &$ 29.1 _{ -3.6}^{+ 3.7 }$ &$ 0.73 _{ -0.13}^{+ 0.11 }$ &$ 12.4 _{ -0.6}^{+ 0.6}$\\
& 2.3 & 1.00 & $ 4.3 _{ -2.0}^{+ 3.2 }$ &$ 28.2 _{ -3.3}^{+ 3.2 }$ &$ 0.74 _{ -0.12}^{+ 0.10 }$ &$ 12.2 _{ -0.1}^{+ 0.8}$\\

\end{longtable}
\end{small}
\end{center}


\bsp	
\label{lastpage}
\end{document}